\documentclass[reprint, superscriptaddress, preprintnumbers,aps,pra]{revtex4-2}
\usepackage{amsmath}
\usepackage{bm}
\usepackage{graphicx}
\usepackage{minted}
\usepackage{mathrsfs}
\usepackage{amssymb}
\usepackage{physics}
\usepackage{braket}
\usepackage{multirow}
\usepackage[normalem]{ulem}
\usepackage{booktabs, ctable}
\usepackage{xcolor, color, framed}
\usepackage[colorlinks, 
linkcolor=red,anchorcolor=green,citecolor=blue]{hyperref}
\usepackage{ulem}
\usepackage{mathtools}
\usepackage{cprotect}
\usepackage{qcircuit}

\renewcommand{\Tr}{\text{Tr}}
\renewcommand{\tr}{\text{tr}}
\newcommand{\Ntr}{N_O^{\mathrm{tr}}}

\newcommand{\PN}{\boldsymbol{P}_N}

\newcommand{\PNtr}{\boldsymbol{P}_N^{\text{tr}}}

\newcommand{\bos}{\boldsymbol}
\newcommand{\OO}{\mathcal{O}}
\definecolor{darkgreen}{rgb}{0.0,0.45,0.0}

\begin{document}
\preprint{RIKEN-iTHEMS-Report-26}

\title{Learning Quantum Operator Dynamics from Short-Time Data}

\author{Jinyang Li}
\email{jinyang.li@riken.jp}
\affiliation{RIKEN Center for Interdisciplinary Theoretical and Mathematical Sciences (iTHEMS), Wako, Saitama 351-0198, Japan}
\affiliation{KEK Theory Center, Institute of Particle and Nuclear Studies}
\affiliation{Graduate University for Advanced Studies (SOKENDAI), Oho 1-1, Tsukuba, Ibaraki 305-0801, Japan}

\author{Satoshi Iso}
\email{satoshi.iso@riken.jp}
\affiliation{RIKEN Center for Interdisciplinary Theoretical and Mathematical Sciences (iTHEMS), Wako, Saitama 351-0198, Japan}
\affiliation{KEK Theory Center, Institute of Particle and Nuclear Studies}

\author{Shunji Matsuura}
\email{shunji.matsuura@riken.jp}
\affiliation{RIKEN Center for Interdisciplinary Theoretical and Mathematical Sciences (iTHEMS), Wako, Saitama 351-0198, Japan}
\affiliation{Department of Electrical and Computer Engineering,
University of British Columbia, Vancouver, BC V6T 1Z4, Canada}
\affiliation{Center for Mathematical Science and Advanced Technology,
Japan Agency for Marine-Earth Science and Technology,
Yokohama 236-0001, Japan}
\affiliation{Department of Physics, University of Guelph, ON N1G 1Y2, Canada}

\author{Lingxiao Wang}
\email{lingxiao.wang@riken.jp}
\affiliation{RIKEN Center for Interdisciplinary Theoretical and Mathematical Sciences (iTHEMS), Wako, Saitama 351-0198, Japan}
\affiliation{Institute for Physics of Intelligence, Graduate School of Science, The University of Tokyo, Bunkyo-ku, Tokyo 113-0033, Japan}

\author{Xiaoyang Wang}
\email{xiaoyang.wang@riken.jp}
\affiliation{RIKEN Center for Interdisciplinary Theoretical and Mathematical Sciences (iTHEMS), Wako, Saitama 351-0198, Japan}
\affiliation{RIKEN Center for Computational Science (R-CCS), Kobe 650-0047, Japan}

\date{\today}

\begin{abstract}
Real-time dynamics of quantum observables provide direct access to excitation spectra and correlation functions in quantum many-body systems, but currently available quantum devices are limited to short evolution times due to decoherence. We propose a neural ordinary differential equation (Neural ODE) framework with a physics-driven neural network to reconstruct long-time operator dynamics from short-time quantum data. By expanding observables in the Pauli basis and exploiting locality and symmetry constraints, the operator evolution is reduced to a tractable set of coefficients whose dynamics are learned from the measurement data. Applied to the Ising spin chain with mixed fields, the operator learning method accurately extrapolates its long-time correlation function and resolves its excitation spectrum. We validate the framework by collecting short-time data from IBM's superconducting quantum processor and successfully reconstruct the Fourier spectrum of an $N=8$ Ising spin chain with 5 times enhanced precision. Our results demonstrate a scalable and data-efficient strategy for extracting dynamical and spectral information from practical quantum hardware.
\end{abstract}

\maketitle

\noindent \emph{Introduction}~---~Real-time dynamics of quantum operators play a central role in characterizing quantum many-body systems across condensed matter physics, high-energy physics, and quantum chemistry. In the Heisenberg picture, the time evolution of observables encodes the excitation spectrum, correlation functions, and transport properties of the system. Simulating such dynamics is computationally challenging on classical computers, while quantum devices provide a natural platform for implementing real-time evolution and measuring observables directly~\cite{Daley_22, King:2025mgn, shinjo2024unveilingcleantwodimensionaldiscrete,yoshioka2024hunting,Neill:2021wla,Efekan_2024,JinZhaoSun2025}. Consequently, dynamical measurements are widely regarded as one of the most promising routes toward demonstrating practical quantum advantage. However, on current noisy quantum hardware, the accessible evolution time is severely limited by decoherence. The restriction to short-time measurements leads to poor frequency resolution in dynamical observables due to the time–frequency uncertainty relation, thereby obscuring key physical information such as excitation spectra and energy gaps~\cite{Neill:2021wla,Efekan_2024,JinZhaoSun2025,Ghim:2024pxe,zhai2025universalquantumcomputationalspectroscopy,z126-zdqj}.


A natural question is therefore whether the intrinsic long-time dynamics of a quantum system can be reconstructed from short-time data. Recently, machine learning approaches have shown promising capabilities in modeling complex dynamical systems and extrapolating temporal behavior from high-dimensional data~\cite{yu2024learning,gajamannage2023recurrent}. In particular, physics-informed learning frameworks incorporate physical priors such as symmetries and conservation laws to improve generalization and interpretability~\cite{yu2024learning,watson2024machine,Aarts:2025gyp}. These ideas have been successfully applied to learning physical dynamics in classical and quantum systems, including governing equation discovery and Hamiltonian and state reconstruction~\cite{greydanus2019hamiltonian,han2021adaptable,chen2021physics,Wang:2025pbd,Raissi:2017zsi,karniadakis2021physics,Liu:2025fix}. In the context of quantum many-body systems, machine learning techniques have been used to learn quantum states~~\cite{Torlai:2020nyq}, reconstruct density matrices~\cite{Lohani:2020slt}, and infer dynamics from measurement data~\cite{Qi:2025zdn,Shah:2024hfe,An:2024yfx,han2021tomography,Huang:2022sqz,Choi:2021uwz}, highlighting the potential of data-driven approaches for quantum system characterization~\cite{Gebhart:2022mhu}.

\begin{figure*}[hbtp!]
    \centering
    \includegraphics[width=0.86\textwidth]{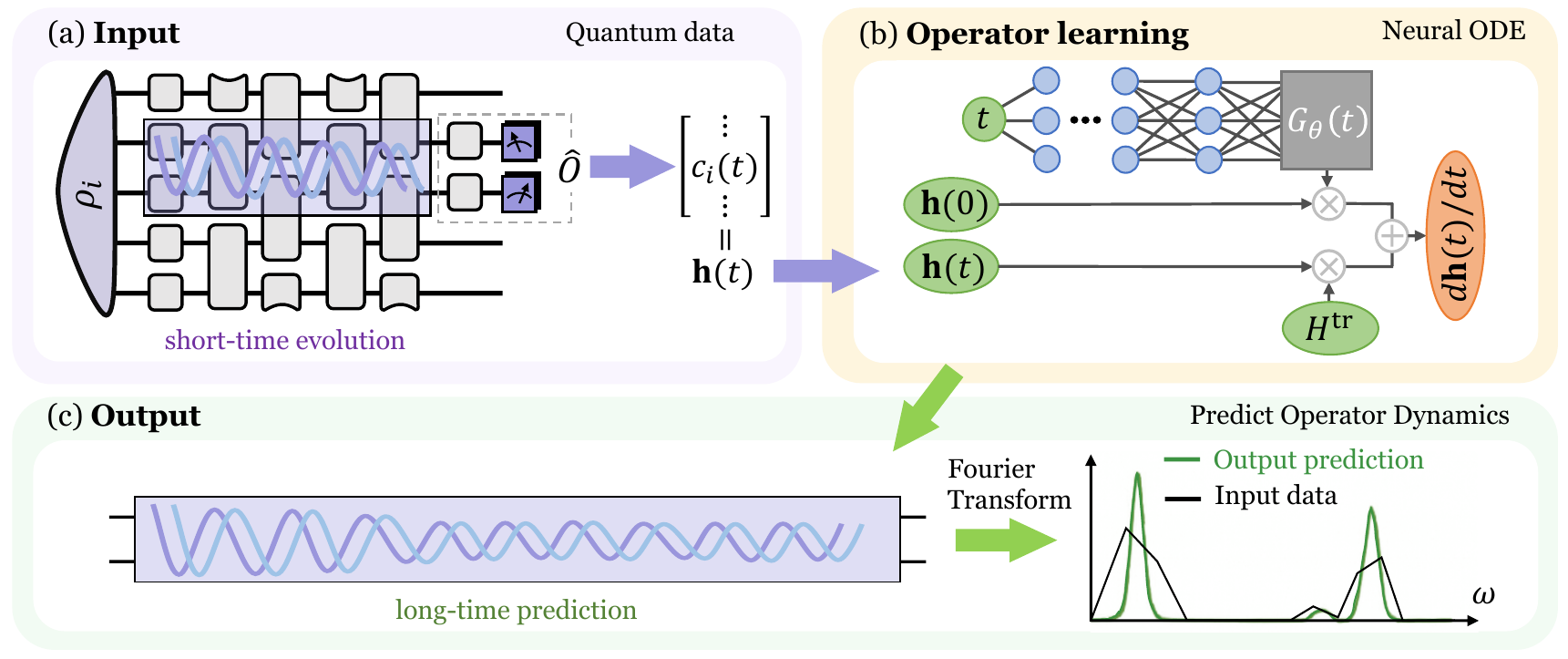}
    \caption{Schematic illustration of the operator-learning framework. (a)~\textbf{Input:} Short-time data of Pauli coefficients $c_i(t)$ to be measured by the evolution of the initial state $\rho_i$ on quantum hardware. (b)~\textbf{Operator learning:} The operator dynamics is learned using the short-time data and a \textit{physics-driven} designed Neural ODE encoding the evolution equation of the truncated coefficients $\mathbf{h}(t)=(c_1(t),\ldots, c_{\Ntr}(t))^T$. (c)~\textbf{Output:} The trained network outputs $d\mathbf{h}(t)/dt$ at an arbitrary $t$, enabling the long-time prediction of $c_i(t)$ and $O(t)$ (left panel). The long-time prediction corresponds to enhanced spectral resolution compared with the input data. (right panel)}
    \label{fig:flowchart}
\end{figure*}

In this Letter, we develop an operator-learning framework that reconstructs long-time quantum dynamics from short-time measurement data. Our approach focuses on the Heisenberg evolution of observables, which naturally encodes the physical information of interest. By exploiting locality and symmetry constraints, we reduce the exponentially large operator space to a tractable set of local Pauli coefficients, ensuring the scalability of the operator learning method. We then design a frequency-aware network (FAN) to learn the evolution of these reduced Pauli coefficients using the neural ordinary differential equations (Neural ODEs)~\cite{chen2018neuralode}, whose continuous-time formulation naturally reflects the differential structure of quantum dynamics. Compared with Hamiltonian-learning approaches~\cite{Hangleiter_2024,PhysRevLett.130.200403}, learning operator dynamics offers several advantages: the dynamics are restricted to energy and symmetry sectors relevant to the observable, the oscillatory structure of operator evolution is robust to low-frequency hardware noise~\cite{Neill:2021wla, Efekan_2024,JinZhaoSun2025,zhai2025universalquantumcomputationalspectroscopy}, and the learned dynamics directly yield physically relevant correlation functions~\cite{z126-zdqj,PhysRevA.65.042323,PhysRevLett.113.020505,Lorenzo_2024,Roggero_2019,Kosugi_2020,Ciavarella_2020, Chen_2021}. Using the Ising spin chain with mixed fields as a benchmark, we show that the proposed framework can accurately reconstruct long-time operator dynamics from short-time data, and enhance the precision and accuracy of the excitation spectrum even in the presence of realistic hardware noise.

\vspace{1.5mm}
\noindent \emph{Operator Dynamics}~---~Consider a quantum system on a $D$-dimensional lattice described by a $k$-local Hamiltonian $H$~\cite{Lloyd:1996aai,Kempe:2004sak}. We aim to predict the long-time operator expectation value $\mathrm{Tr}[O(t)\rho_0]$ for an arbitrary initial state $\rho_0$ and a Heisenberg-evolved observable $O(t) = e^{iHt}Oe^{-iHt}$. This time-evolved operator can be decomposed in a complete $N$-qubit Pauli string basis $\bos{P}_{N}:=\{I,X,Y,Z\}^{\otimes N}$
\begin{align}
    O(t) = \sum_{\sigma_i \in \bos{P}_{N}}c_i(t) \sigma_i,
    \label{eq:Ot-expansion}
\end{align}
where $c_i(t)$ are real coefficients due to the Hermiticity of the observable $O$. Since the Pauli string expectation $\Tr(\sigma_i \rho_0)$ can be evaluated if $\rho_0$ can be prepared on quantum computers, or using classical tensor network and Monte-Carlo methods~\cite{ORUS2014117,gubernatis2016quantum,Huggins2022,jt8s-hzhd,Berezutskii_25}, predicting the long-time behavior of $\mathrm{Tr}[O(t)\rho_0]$ is converted to predicting the long-time evolution of $c_i(t)$.

The long-time evolution of $c_i(t)$ can be learned by firstly acquiring the short-time data from quantum computers. For a given $\sigma_i$, its coefficient $c_i(t)$ is measured via the formula $c_i(t)=\mathrm{Tr}\!\left[O(t)\,\sigma_i\right]/d$, which is derived from Eq.~\eqref{eq:Ot-expansion} according to the orthogonality of the Pauli basis. Here $d=2^N$ is the dimension of the Hilbert space. We assume that $O$ can be written as a linear combination of local Pauli strings without identity, which is satisfied in many physically relevant applications. Thus, $O$ is traceless, and $c_i(t)$ can be measured by (i)preparing an initial state with density matrix $\rho_i:=(\sigma_i+\mathbb{I})/d,~\mathbb{I}:=I^{\otimes N}$ using the quantum circuit given in End Note; (ii)evolving the state by quantum circuits using, e.g., Trotter decomposition~\cite{SUZUKI1990319}; and finally (iii)measuring the expectation of the observable $O$. These data acquisition procedures are illustrated in Fig.~\ref{fig:flowchart}(a). 

Measuring all $c_i(t)$s of the complete Pauli basis $\bos{P}_N$ requires quantum and classical time complexity growing exponentially with the system size $N$. To deal with this issue, we introduce a local Pauli truncation, in which only coefficients of spatially local and symmetry-preserving Pauli strings are measured. After the truncation, we prove that the number of measured Pauli coefficients $c_i(t)$ scales as $N_O^{\mathrm{tr}}\sim N4^{\OO(kT^D)}$, where $T$ is the short evolution time~\cite{supp}. This linear dependence on $N$ ensures the scalability of the data acquisition process of the operator learning method. In practice, the training data time should be longer than $T$ to ensure adequate neural network training~\cite{supp}.

We then predict the long-time evolution of $c_i(t)$ for these locally truncated Pauli strings $\sigma_i\in \PN^{\tr}$. The resulting operator $O_{\tr}(t) = \sum_{\sigma_i\in \PN^{\tr}}c_i(t)\sigma_i$ resembles the original $O(t)$ if $O(t)$ is spatially localized. For non-localized $O(t)$, the spectral information of $O(t)$ is still contained in the local Pauli coefficients $c_i(t)$, such that we can resolve the spectrum of $O(t)$ using $O_{\tr}(t)$. More detailed argument by the spectral decomposition of $O(t)$, $c_i(t)$, and the numerical studies of the long-time truncation error are given in Supplemental Material~\cite{supp}.

The truncated Pauli coefficients are time-evolved by an ordinary differential equation
\begin{align}
    \frac{d\mathbf{h}(t)}{dt}=H^{\tr}\mathbf{h}(t)+G(t)\mathbf{h}(0),
    \label{eq:truncated-ODE-2}
\end{align}
where the local coefficients $\mathbf{h}(t):=(c_1(t),\ldots,c_{N_O^{\mathrm{tr}}}(t))^T$. Both $H^{\tr}$ and $G(t)$ are $N_O^{\mathrm{tr}}\times N_O^{\mathrm{tr}}$ matrices: $H^{\tr}_{ij}:=-i\Tr(H[\sigma_i,\sigma_j])/d$ is the Hamiltonian in the truncated local Pauli basis; $G(t)$ is a propagator matrix relating the truncated non-local Pauli strings $\sigma_i\in \PN/ \PN^{\tr}$ to the local ones in $\PN^{\tr}$. The explicit form of $G(t)$ and the derivation of Eq.~\eqref{eq:truncated-ODE-2} are given in End Note. If $G(t)$ can be explicitly solved, the trajectories of $\mathbf{h}(t)$ are obtained from $d\mathbf{h}(t)/dt$ using a suitable update rule, such as the Runge-Kutta method. However, $G(t)$ is hard to solve due to contributions from exponentially many non-local Pauli basis elements. Instead, in the next section, we show that $G(t)$ can be learned using short-time data and deep neural networks. 

\vspace{1.5mm}
\noindent \emph{Neural Operator Learning}~---~According to the universal approximation theorem of deep neural networks~\cite{Cybenko1989,FUNAHASHI1989183}, we can in principle find a deep neural network to represent the propagator matrix $G(t)$, such that
\begin{align}
\frac{d\mathbf{h}(t)}{dt}=H^{\tr}\mathbf{h}(t)+G_{\theta}(t)\mathbf{h}(0),
\label{eq:ODE-neural}
\end{align}
where $G_{\theta}(t)$ is the $N_O^{\mathrm{tr}}\times N_O^{\mathrm{tr}}$-dimensional output of a deep neural network with trainable parameters $\{\mathbf{\theta}\}$. These parameters are trained using the Neural ODE illustrated in Fig.~\ref{fig:flowchart}(b). This continuous-time formulation learns smooth, physically consistent trajectories and naturally reflects the differential structure of quantum dynamics. However, if we require Eq.~\eqref{eq:ODE-neural} to hold exactly, $G_{\theta}(t)$ may have network layers or hidden dimensions grow exponentially with the input dimension $N_O^{\mathrm{tr}}$.

Instead, we can use an $L$-layer fully connected network (FCN) to approximate $G_{\theta}(t)$~\cite{chen2018neuralode,bishop2023deep}, i.e., $\hat{G}_{\theta}(t)=\mathbf{W}_L\!\left(\phi_{L-1}\circ\phi_{L-2}\circ\cdots\circ\phi_{0}\right)(t)+\mathbf{b}_L$, and the $l$-th layer output is $\phi_l(\mathbf{x}_l)\coloneqq\tanh(\mathbf{W}_l\mathbf{x}_l+\mathbf{b}_l)$. The dimension of the trainable weights $\mathbf{W}_l$ and biases $\mathbf{b}_l$ match the dimension of the input $\mathbf{x}_l$ and the output $\mathbf{x}_{l+1}$. However, such a naive architecture is not only computationally expensive but also prone to overfitting, especially when the training data is limited.

\begin{figure}
    \centering
    \includegraphics[width=0.45\textwidth]{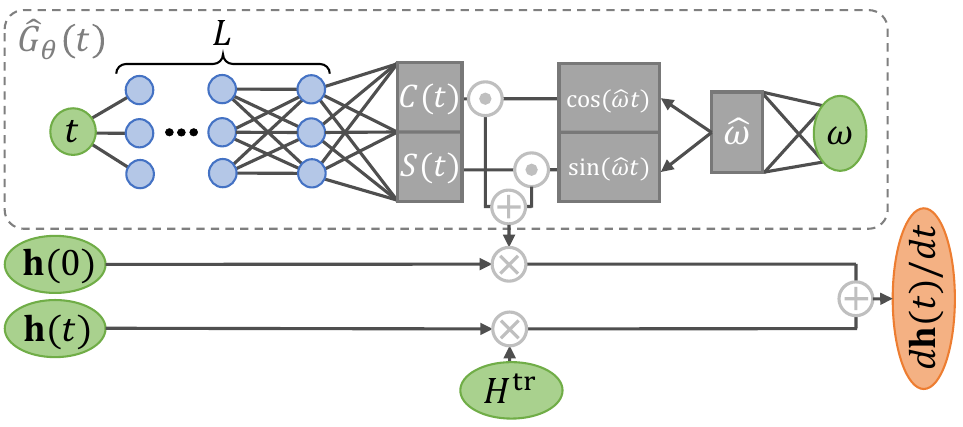} 
    \caption{Frequency-aware network architecture. Pauli coefficients $\mathbf{h}(t)$, time $t$, local Pauli truncated Hamiltonian $H^{\tr}$, and raw frequencies $\omega$ in the green circles are the input and $d\mathbf{h}(t)/dt$ in the orange circle is the output. The circled dot, plus, and cross denote the inner product, addition, and matrix multiplication, respectively.} 
    \label{fig:FAN}
\end{figure}

To improve the ability of FCNs for long-time prediction, we adopt a \textit{physics-driven learning} design~\cite{Aarts:2025gyp} to systematically incorporate essential physical priors into deep learning models. In End Note, we show that $G(t)$ contains many periodic modes, making frequency-aware modules a promising choice for the latent layers. Compared to FCN, periodic modules offer a more efficient representation of oscillatory functions by mitigating the inherent spectral bias of conventional networks~\cite{sitzmann2019siren,tancik2020fourfeat}. Based on these insights into sinusoidal feature efficiency~\cite{Marieme2021fnn}, we introduce a frequency-aware network (FAN) designed as follows,
\begin{align}
(\hat{G}_{\theta})_{ij} &= \sum_{m=1}^{N_f}\left[C_{ijm}(t)\cos(\hat{\omega}_mt)+S_{ijm}(t)\sin(\hat{\omega}_m t)\right],\nonumber
\end{align}
where $N_f$ is the total number of frequencies expected to be learned from the data. We include both cosine and sine terms to effectively account for the initial phases of the periodic modes. The envelope function $C_{ijm}(t),S_{ijm}(t)$ and frequency $\hat{\omega}_m$ are parametrized as 
\begin{equation}
    \begin{aligned} 
\begin{bmatrix}
C_{ijm}\\
S_{ijm}
\end{bmatrix}&=\mathbf{W}^{(e)}_L(\phi_{L-1}\circ\phi_{L-2}\circ\cdots \circ \phi_0)(t)+\mathbf{b}_{L}^{(e)};\\
\hat{\omega}_m &= \sum_{m'=1}^{N_f'}\mathbf{W}^{(f)}_{mm'}\omega_{m'} .
\end{aligned}
\end{equation}
Here, $\phi_i$ indicates a fully connected layer with the $\tanh$ activation function, and we set their dimensions uniformly to a hidden dimension $d_h$. $\hat{\omega}_m$ is a linear transformation of raw frequencies $\omega_m$ with learnable parameters $\mathbf{W}^{(f)}$. Raw frequencies $\omega_m$ are the physically relevant frequencies obtained from the Fourier transform of raw training data, which will be detailed in the following numerical simulations. Combined with Eq.~\eqref{eq:ODE-neural}, the full FAN is schematically shown in Fig.~\ref{fig:FAN}. For fixed network layers $L$ and hidden dimension $d_h$, the number of network parameters is of $\OO((\Ntr)^2)$, ensuring the scalability of the FAN training using Neural ODE.
\vspace{1.5mm}

\begin{figure}
    \centering
    \includegraphics[width=0.48\textwidth]{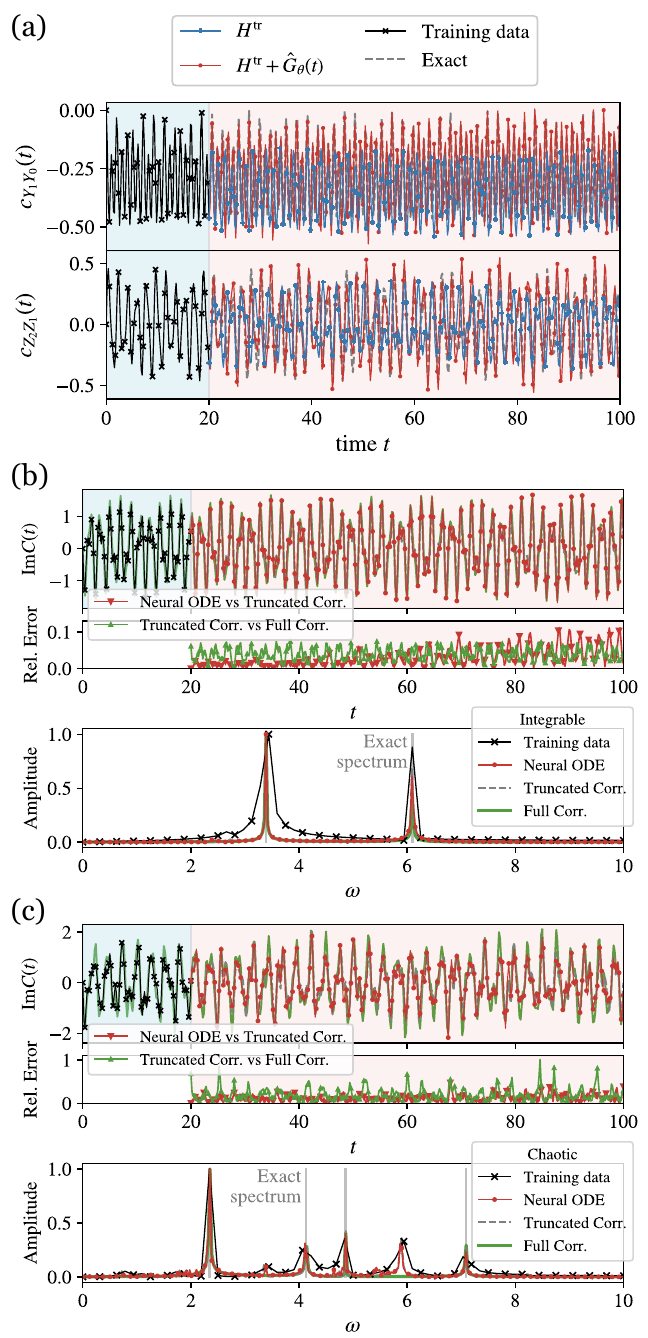} 
    \caption{(a) Long-time prediction of the Pauli coefficients $c_i(t)$ versus time $t$ in $N=3$ transverse-field Ising model (TFIM) with local Pauli truncation $w\leq 2$.  The short-time training data is $t\leq 20$ with a blue background. The ground-truth results, predictions using the truncated Hamiltonian $H^{\tr}$, and $H^{\tr}$ plus FAN $\hat{G}_{\theta}(t)$ are plotted using black-cross, blue, and red-dotted lines, respectively. Imaginary part of the correlation function versus time $t$ for (b) the integrable TFIM and (c) the chaotic model. Results obtained from Neural ODE prediction, $w\leq 2$ truncated Pauli basis, and full Pauli basis are denoted by red-dotted, gray-dashed, and green-solid lines, respectively. The relative errors between these three lines are given in the middle panel, and the bottom panel plots the spectrum obtained by the Fourier transform of $\text{Im} C(t)$.}
    \label{fig:numerics}
\end{figure}

\noindent \emph{Numerical results}~---~To benchmark our operator-learning method, we perform numerical simulations and hardware demonstrations for the Ising spin chain with mixed fields. The Hamiltonian reads $H=-\sum_{i=1}^{N-1} Z_iZ_{i+1}-h\sum_{i=1}^{N} X_i-m\sum_{i=1}^N Z_i$, where $Z_i, X_i$ are Pauli matrices on the $i$-th site. We consider two sets of parameters: (i) $h=1, m=0$ is the integrable transverse-field Ising model (TFIM), and (ii) $h=-1.05, m=0.5$ is known to be a chaotic system~\cite{PhysRevLett.106.050405,PhysRevD.104.074518}. We aim to predict the long-time behavior of the two-point correlation function
\begin{equation}
C(t):= \bra{\Omega}  O_X(t) O_X(0) \ket{\Omega},
\label{Order_3qub}
\end{equation}
where $\ket{\Omega}$ is the ground state of the Hamiltonian. The operator $O_X \coloneqq \sum_{i=1}^N X_i$ excites a pair of domain walls with zero center-of-mass momentum from the vacuum state~\cite{Zhang2017,PhysRevLett.122.150601,PRXQuantum.3.040309}. The long-time evolution of $C(t)$ gives higher resolution of the domain wall spectrum according to the uncertainty principle $\delta \omega\sim 2\pi/T$. $C(t)$ encodes the sparse low-energy excitation spectrum of the Hamiltonian system, and this spectral sparsity allows the neural network to learn it efficiently. In the numerical simulation below, we use the frequencies extracted from the spectrum of $C(t)$ over the short-time training window as the input frequencies $\omega_m$ supplied to the neural network, and the target number of frequencies equals the input ones.

We first compare the operator dynamics with and without the propagator matrix learned by FAN. In the TFIM with $N=3$, Fig.~\ref{fig:numerics}(a) shows the predicted long-time evolution of Pauli coefficients $c_i(t)$ with locally truncated Pauli weight $w\leq 2$. We present two coefficients $c_{i}(t)$ of $\sigma_i=Y_1Y_0$ and $Z_2Z_1$ among $11$ spatially local and symmetry-preserving Pauli bases. The FAN is trained on short-time data $t\in[0,20]$ from exact diagonalization with step length $ \delta t=0.2$, using the envelope layers $L=3$ and the hidden dimension $d_h=10$. Then, the Neural ODE extrapolates the operator dynamics up to $t=100$. We see that for all coefficients $c_i(t)$, the predicted trajectories by $H^{\tr}$ plus FAN $\hat{G}_{\theta}$ are well consistent with the results by exact diagonalization. On the other hand, coefficients evolution by only the truncated Hamiltonian $H^{\tr}$ deviates from the exact one, showing that the propagator matrix is necessary for accurately predicting the long-time operator dynamics, and the matrix can be well learned by FAN using the short-time data. In Supplemental Material~\cite{supp}, we perform additional ablation studies to investigate the necessity of long enough training data time, FAN comparison with FCN, the envelope network with a large enough $L$, and the symmetry constraints on $c_i(t)$.

Using the predicted $c_i(t)$, we reconstruct the correlation function $C(t)$ of the integrable TFIM and the chaotic model shown in Fig.~\ref{fig:numerics}(b) and (c), respectively. Compared with the full correlation function, two dominant errors of the operator learning method are from the local Pauli truncation and the Neural ODE prediction, which are respectively reflected by the relative error between the truncated correlations versus the full correlations, and the Neural ODE results versus the truncated correlations plotted in the middle panel of Fig.~\ref{fig:numerics}(b,c). For the integrable TFIM, we see that both relative errors are smaller than $0.1$ for $t\leq 100$, whereas for the chaotic model, the Pauli truncation error is significantly larger than the still-small prediction error. This is reasonable because the chaotic model usually have $O_X(t)$ fast spreading across the whole system~\cite{Swingle2018,PhysRevLett.129.160602,liang2024observationanomalousinformationscrambling,Abanin2025}, making $O_{X,\tr}(t)$ deviate from $O_{X}(t)$. Additionally, the prediction error can also be large in the case of strong spatial coupling and dense excitation spectrum~\cite{supp}. However, despite these errors in the time domain, the spectrum obtained from the Fourier transform of $C(t)$ remains well consistent with the exact spectrum in the lower panel of Fig.~\ref{fig:numerics}(b,c). Moreover, the spectrum obtained from Neural ODE shows a clear improvement in precision and accuracy over those extracted from the training data.

\begin{figure}
\centering
\includegraphics[width=0.48\textwidth]{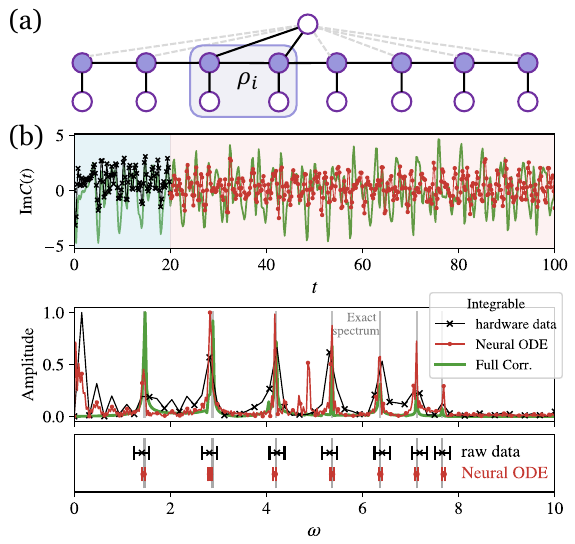}
\cprotect\caption{(a) Connection of the $N=8$ TFIM system qubits (filled circles) and the ancillary qubits (empty circles) on the \verb|ibm_phoenix| to measure the short-time data $c_i(t)=\mathrm{Tr}(O_X(t)\rho_i)$. Solid lines denote the two-qubit gate connection. (b) Imaginary part of the correlation function in the time domain (upper panel) and in the frequency domain (lower panel). Black-cross and red-dotted lines are respectively obtained from the measured data $c_i(t)$ with $t\in[0,20]$ and Neural ODE prediction for $t\in(20,100]$. Gray solid lines are the spectrum from exact diagonalization.}
\label{fig:hardware}
\end{figure}

Finally, we implement the operator learning method on the 120-qubit \verb|ibm_phoenix| quantum hardware~\cite{Qiskit}, where $8$ system qubits are used to represent an $N=8$ TFIM spin system, and $9$ ancillary qubits are used to prepare the initial state $\rho_i$ (See End Note), as respectively denoted by the filled and empty circles in Fig.~\ref{fig:hardware}(a). The measured training data $c_{i}(t)$ of $36$ Pauli bases with local weight $w\leq 2$ and hardware implementation details are given in Supplemental Material~\cite{supp}. Here, $T/\delta t=100$ steps of first-order Trotter evolution circuits are implemented and compressed using the algebraic compression method~\cite{PhysRevA.105.032420}. The resulting short-time correlation function, the predicted results by Neural ODE, and the full correlation function are denoted by black-cross, red-dotted, and green lines in Fig.~\ref{fig:hardware}(b). We observe that the quantum noise, Trotter error, truncation error, and prediction error lead to substantial deviations of the Neural ODE results from the full correlation function. Nevertheless, as shown in the lower panel of Fig.~\ref{fig:hardware}(b), the Neural ODE can still learn from the noisy short-time data and reconstruct a Fourier spectrum that is $5$ times more accurate than that obtained from the short-time data according to $\delta\omega\sim 2\pi/T$. The reconstructed spectral peaks further agree with the exact diagonalization spectrum (gray solid lines) within the energy uncertainty. These results demonstrate the applicability of the operator learning method in the presence of Trotter error and hardware noise.
\vspace{1.5mm}

\noindent \emph{Summary}~---~In this Letter, we present an operator learning framework to predict long-time operator dynamics from short-time quantum data. By imposing the local Pauli truncation, the exponentially large operator space is reduced to a tractable manifold that preserves the essential spectral information of the operator dynamics. Then, the FAN is designed to learn and predict the Pauli evolution in the truncated manifold, thereby improving spectral resolution beyond the limited coherence time imposed by decoherence and hardware noise. We demonstrate the method on the Ising spin chain with mixed fields to reconstruct the long-time correlation function and improve the model's spectral resolution. Its applicability is further demonstrated using IBM's superconducting quantum processor in the presence of Trotter error and hardware noise. Beyond predicting observables, the ability of the framework to learn effective operator evolution under noise suggests potential applications to quantum error correction (QEC), such as noise-aware decoding and the characterization of time-dependent correlated errors~\cite{Baireuther:2018mlm,Flurin:2018rzr,Valenti:2019ohp}. Overall, our results establish Neural ODEs as a scalable, hardware-compatible tool for Hamiltonian learning and quantum system identification on practical quantum hardware.
\vspace{1.5mm}

\noindent \emph{Acknowledgment}~---~We thank Drs.\ Tetsuo Hatsuda, Xingyu Guo, Keren Li, and Shuzhe Shi for helpful discussions.
We thank the DEEP-IN working group at RIKEN-iTHEMS for support in the preparation of this paper.
L.W. is supported by JSPS KAKENHI Grant No. 25H01560, and JST-BOOST Grant No. JPMJBY24H9.
X.W. is supported by the RIKEN TRIP initiative (RIKEN Quantum) and the UTokyo Quantum Initiative.

\nocite{Lieb:1972wy, PhysRevLett.97.050401,Dinh:2016pgf}

\bibliography{ref}

@misc{supp,
howpublished = {See Supplemental Material at \url{Supplemental/Material/website} for the operator truncation and symmetry reduction scheme, prediction error and ablation studies regarding the FAN network, different data sizes and external field strength, and the symmetry constraints, properties of the two-point correlation function, and details of the hardware implementation and Neural ODE training, with additionally included Refs. [75-80]}
}

@article{Choi:2021uwz,
    author = "Choi, Matthew and Flam-Shepherd, Daniel and Kyaw, Thi Ha and Aspuru-Guzik, Al{\'a}n",
    title = "{Learning quantum dynamics with latent neural ordinary differential equations~}",
    eprint = "2110.10721",
    archivePrefix = "arXiv",
    primaryClass = "quant-ph",
    doi = "10.1103/PhysRevA.105.042403",
    journal = "Phys. Rev. A",
    volume = "105",
    number = "4",
    pages = "042403",
    year = "2022"
}

@article{Lieb:1972wy,
    author = "Lieb, E. H. and Robinson, D. W.",
    title = "{The finite group velocity of quantum spin systems}",
    doi = "10.1007/BF01645779",
    journal = "Commun. Math. Phys.",
    volume = "28",
    pages = "251--257",
    year = "1972"
}

@article{PhysRevLett.97.050401,
  title = {Lieb-Robinson Bounds and the Generation of Correlations and Topological Quantum Order},
  author = {Bravyi, S. and Hastings, M. B. and Verstraete, F.},
  journal = {Phys. Rev. Lett.},
  volume = {97},
  issue = {5},
  pages = {050401},
  numpages = {4},
  year = {2006},
  month = {Jul},
  publisher = {American Physical Society},
  doi = {10.1103/PhysRevLett.97.050401},
  url = {https://link.aps.org/doi/10.1103/PhysRevLett.97.050401}
}

@article{White2009,
  title = {Minimally Entangled Typical Quantum States at Finite Temperature},
  author = {White, Steven R.},
  journal = {Phys. Rev. Lett.},
  volume = {102},
  issue = {19},
  pages = {190601},
  numpages = {4},
  year = {2009},
  month = {May},
  publisher = {American Physical Society},
  doi = {10.1103/PhysRevLett.102.190601},
  url = {https://link.aps.org/doi/10.1103/PhysRevLett.102.190601}
}

@article{Baireuther:2018mlm,
    author = "Baireuther, Paul and O'Brien, Thomas E. and Tarasinski, Brian and Beenakker, Carlo W. J.",
    title = "{Machine-learning-assisted correction of correlated qubit errors in a topological code}",
    eprint = "1705.07855",
    archivePrefix = "arXiv",
    primaryClass = "quant-ph",
    doi = "10.22331/q-2018-01-29-48",
    journal = "Quantum",
    volume = "2",
    pages = "48",
    year = "2018"
}

@article{Flurin:2018rzr,
    author = "Flurin, Emmanuel and Martin, Leigh S. and Hacohen-Gourgy, Shay and Siddiqi, Irfan",
    title = "{Using a Recurrent Neural Network to Reconstruct Quantum Dynamics of a Superconducting Qubit from Physical Observations}",
    eprint = "1811.12420",
    archivePrefix = "arXiv",
    primaryClass = "quant-ph",
    doi = "10.1103/PhysRevX.10.011006",
    journal = "Phys. Rev. X",
    volume = "10",
    number = "1",
    pages = "011006",
    year = "2020"
}

@article{Valenti:2019ohp,
    author = "Valenti, Agnes and van Nieuwenburg, Evert and Huber, Sebastian and Greplova, Eliska",
    title = "{Hamiltonian Learning for Quantum Error Correction}",
    eprint = "1907.02540",
    archivePrefix = "arXiv",
    primaryClass = "quant-ph",
    doi = "10.1103/PhysRevResearch.1.033092",
    journal = "Phys. Rev. Res.",
    volume = "1",
    pages = "033092",
    year = "2019"
}

@article{Gebhart:2022mhu,
    author = "Gebhart, Valentin and Santagati, Raffaele and Gentile, Antonio Andrea and Gauger, Erik M. and Craig, David and Ares, Natalia and Banchi, Leonardo and Marquardt, Florian and Pezze', Luca and Bonato, Cristian",
    title = "{Learning quantum systems}",
    eprint = "2207.00298",
    archivePrefix = "arXiv",
    primaryClass = "quant-ph",
    reportNumber = "FERMILAB-PUB-22-937-SQMS-V",
    doi = "10.1038/s42254-022-00552-1",
    journal = "Nature Rev. Phys.",
    volume = "5",
    number = "3",
    pages = "141--156",
    year = "2023"
}

@inproceedings{sitzmann2019siren,
    author = {Sitzmann, Vincent and Martel, Julien N. P. and Bergman, Alexander W. and Lindell, David B. and Wetzstein, Gordon},
    title = {Implicit neural representations with periodic activation functions},
    year = {2020},
    isbn = {9781713829546},
    publisher = {Curran Associates Inc.},
    address = {Red Hook, NY, USA},
    booktitle = {Proceedings of the 34th International Conference on Neural Information Processing Systems},
    articleno = {626},
    numpages = {12},
    location = {Vancouver, BC, Canada},
    series = {NIPS '20}
}

@inproceedings{tancik2020fourfeat,
    title={Fourier Features Let Networks Learn High Frequency Functions in Low Dimensional Domains},
    author = {Tancik, Matthew and Srinivasan, Pratul P. and Mildenhall, Ben and Fridovich-Keil, Sara and Raghavan, Nithin and Singhal, Utkarsh and Ramamoorthi, Ravi and Barron, Jonathan T. and Ng, Ren},
    title = {Fourier features let networks learn high frequency functions in low dimensional domains},
    year = {2020},
    isbn = {9781713829546},
    publisher = {Curran Associates Inc.},
    address = {Red Hook, NY, USA},
    booktitle = {Proceedings of the 34th International Conference on Neural Information Processing Systems},
    articleno = {632},
    numpages = {11},
    location = {Vancouver, BC, Canada},
    series = {NIPS '20}
}

@article{Marieme2021fnn,
    author = {Ngom, Marieme and Marin, Oana},
    title = {Fourier neural networks as function approximators and differential equation solvers},
    year = {2021},
    issue_date = {December 2021},
    publisher = {John Wiley \& Sons, Inc.},
    address = {USA},
    volume = {14},
    number = {6},
    issn = {1932-1864},
    url = {https://doi.org/10.1002/sam.11531},
    doi = {10.1002/sam.11531},
    journal = {Stat. Anal. Data Min.},
    month = nov,
    pages = {647–661},
    numpages = {15}
}

@article{PhysRevLett.130.200403,
  title = {Learning Many-Body Hamiltonians with Heisenberg-Limited Scaling},
  author = {Huang, Hsin-Yuan and Tong, Yu and Fang, Di and Su, Yuan},
  journal = {Phys. Rev. Lett.},
  volume = {130},
  issue = {20},
  pages = {200403},
  numpages = {7},
  year = {2023},
  month = {May},
  publisher = {American Physical Society},
  doi = {10.1103/PhysRevLett.130.200403},
  url = {https://link.aps.org/doi/10.1103/PhysRevLett.130.200403}
}

@article{Hangleiter_2024,
	author = {Hangleiter, Dominik and Roth, Ingo and Fuksa, Jon{\'a}{\v s} and Eisert, Jens and Roushan, Pedram},
	da = {2024/11/06},
	doi = {10.1038/s41467-024-52629-3},
	id = {Hangleiter2024},
	isbn = {2041-1723},
	journal = {Nature Communications},
	number = {1},
	pages = {9595},
	title = {Robustly learning the Hamiltonian dynamics of a superconducting quantum processor},
	ty = {JOUR},
	url = {https://doi.org/10.1038/s41467-024-52629-3},
	volume = {15},
	year = {2024}}

@article{gajamannage2023recurrent,
  title={Recurrent neural networks for dynamical systems: Applications to ordinary differential equations, collective motion, and hydrological modeling},
  author={Gajamannage, K and Jayathilake, DI and Park, Y and Bollt, EM},
  journal={Chaos: An Interdisciplinary Journal of Nonlinear Science},
  volume={33},
  number={1},
  year={2023},
  publisher={AIP Publishing}
}

@article{yu2024learning,
  title={Learning dynamical systems from data: An introduction to physics-guided deep learning},
  author={Yu, Rose and Wang, Rui},
  journal={Proceedings of the National Academy of Sciences},
  volume={121},
  number={27},
  pages={e2311808121},
  year={2024},
  publisher={National Academy of Sciences}
}

@article{watson2024machine,
  title={Machine learning with physics knowledge for prediction: A survey},
  author={Watson, Joe and Song, Chen and Weeger, Oliver and Gruner, Theo and Le, An T and Pompetzki, Kay and Hendawy, Ahmed and Arenz, Oleg and Trojak, Will and Cranmer, Miles and others},
  journal={arXiv preprint arXiv:2408.09840},
  year={2024}
}

@article{Liu:2025fix,
    author = "Liu, Jie and Wang, Xin",
    title = "{Hamiltonian learning via inverse physics-informed neural networks}",
    eprint = "2506.10379",
    archivePrefix = "arXiv",
    primaryClass = "quant-ph",
    doi = "10.1103/nx97-zjdf",
    journal = "Phys. Rev. Res.",
    volume = "7",
    number = "4",
    pages = "043137",
    year = "2025"
}

@article{An:2024yfx,
    author = "An, Zheng and Wu, Jiahui and Lin, Zidong and Yang, Xiaobo and Li, Keren and Zeng, Bei",
    title = "{Dual-Capability Machine Learning Models for Quantum Hamiltonian Parameter Estimation and Dynamics Prediction}",
    eprint = "2405.13582",
    archivePrefix = "arXiv",
    primaryClass = "quant-ph",
    doi = "10.1103/PhysRevLett.134.120202",
    journal = "Phys. Rev. Lett.",
    volume = "134",
    number = "12",
    pages = "120202",
    year = "2025"
}

@article{han2021tomography,
  title={Tomography of time-dependent quantum Hamiltonians with machine learning},
  author={Han, Chen-Di and Glaz, Bryan and Haile, Mulugeta and Lai, Ying-Cheng},
  journal={Physical Review A},
  volume={104},
  number={6},
  pages={062404},
  year={2021},
  publisher={APS}
}

@article{Huang:2022sqz,
    author = "Huang, Hsin-Yuan and Chen, Sitan and Preskill, John",
    title = "{Learning to Predict Arbitrary Quantum Processes}",
    eprint = "2210.14894",
    archivePrefix = "arXiv",
    primaryClass = "quant-ph",
    doi = "10.1103/PRXQuantum.4.040337",
    journal = "PRX Quantum",
    volume = "4",
    number = "4",
    pages = "040337",
    year = "2023"
}

@article{greydanus2019hamiltonian,
  title={Hamiltonian neural networks},
  author={Greydanus, Samuel and Dzamba, Misko and Yosinski, Jason},
  journal={Advances in neural information processing systems},
  volume={32},
  year={2019}
}

@article{han2021adaptable,
  title={Adaptable Hamiltonian neural networks},
  author={Han, Chen-Di and Glaz, Bryan and Haile, Mulugeta and Lai, Ying-Cheng},
  journal={Physical Review Research},
  volume={3},
  number={2},
  pages={023156},
  year={2021},
  publisher={APS}
}

@article{chen2021physics,
  title={Physics-informed learning of governing equations from scarce data},
  author={Chen, Zhao and Liu, Yang and Sun, Hao},
  journal={Nature communications},
  volume={12},
  number={1},
  pages={6136},
  year={2021},
  publisher={Nature Publishing Group UK London}
}

@article{Wang:2025pbd,
  title = {Physics-based discovery of governing equations from scarce and noisy data},
  author = {Wang, Liang and Ji, Tingwei and Huang, Yunzhe and Zhou, Hongjie and Xie, Fangfang},
  journal = {Phys. Rev. E},
  volume = {111},
  issue = {6},
  pages = {065314},
  numpages = {20},
  year = {2025},
  month = {Jun},
  publisher = {American Physical Society},
  doi = {10.1103/d4tm-92vb},
  url = {https://link.aps.org/doi/10.1103/d4tm-92vb}
}

@article{Zhang2017,
	author = {Zhang, J. and Pagano, G. and Hess, P. W. and Kyprianidis, A. and Becker, P. and Kaplan, H. and Gorshkov, A. V. and Gong, Z. -X. and Monroe, C.},
	da = {2017/11/01},
	doi = {10.1038/nature24654},
	id = {Zhang2017},
	isbn = {1476-4687},
	journal = {Nature},
	number = {7682},
	pages = {601--604},
	title = {Observation of a many-body dynamical phase transition with a 53-qubit quantum simulator},
	ty = {JOUR},
	url = {https://doi.org/10.1038/nature24654},
	volume = {551},
	year = {2017}}

@article{Raissi:2017zsi,
    author = "Raissi, Maziar and Perdikaris, Paris and Karniadakis, George Em",
    title = "{Physics-informed neural networks: A deep learning framework for solving forward and inverse problems involving nonlinear partial differential equations}",
    eprint = "1711.10561",
    archivePrefix = "arXiv",
    primaryClass = "cs.AI",
    doi = "10.1016/j.jcp.2018.10.045",
    journal = "J. Comput. Phys.",
    volume = "378",
    pages = "686--707",
    year = "2019"
}

@article{karniadakis2021physics,
  title={Physics-informed machine learning},
  author={Karniadakis, George Em and Kevrekidis, Ioannis G and Lu, Lu and Perdikaris, Paris and Wang, Sifan and Yang, Liu},
  journal={Nature Reviews Physics},
  volume={3},
  number={6},
  pages={422--440},
  year={2021},
  publisher={Nature Publishing Group UK London}
}

@article{Aarts:2025gyp,
    author = "Aarts, Gert and Fukushima, Kenji and Hatsuda, Tetsuo and Ipp, Andreas and Shi, Shuzhe and Wang, Lingxiao and Zhou, Kai",
    title = "{Physics-driven learning for inverse problems in quantum chromodynamics}",
    eprint = "2501.05580",
    archivePrefix = "arXiv",
    primaryClass = "hep-lat",
    reportNumber = "RIKEN-iTHEMS-Report-25",
    doi = "10.1038/s42254-024-00798-x",
    journal = "Nature Rev. Phys.",
    volume = "7",
    number = "3",
    pages = "154--163",
    year = "2025"
}

@preprint{Qi:2025zdn,
    author = "Qi, Zihao and Peng, Yang and Earls, Christopher",
    title = "{Fourier Neural Operators for Time-Periodic Quantum Systems: Learning Floquet Hamiltonians, Observable Dynamics, and Operator Growth}",
    eprint = "2509.07084",
    archivePrefix = "arXiv",
    primaryClass = "quant-ph",
    month = "9",
    year = "2025"
}

@preprint{Shah:2024hfe,
    author = "Shah, Freya and Patti, Taylor L. and Berner, Julius and Tolooshams, Bahareh and Kossaifi, Jean and Anandkumar, Anima",
    title = "{Fourier Neural Operators for Learning Dynamics in Quantum Spin Systems}",
    eprint = "2409.03302",
    archivePrefix = "arXiv",
    primaryClass = "quant-ph",
    month = "9",
    year = "2024"
}

@article{Lohani:2020slt,
    author = "Lohani, Sanjaya and Kirby, Brian T. and Brodsky, Michael and Danaci, Onur and Glasser, Ryan T.",
    title = "{Machine learning assisted quantum state estimation}",
    doi = "10.1088/2632-2153/ab9a21",
    journal = "Mach. Learn. Sci. Tech.",
    volume = "1",
    number = "3",
    pages = "035007",
    year = "2020"
}

@article{Torlai:2020nyq,
    author = "Torlai, Giacomo and Melko, Roger G.",
    title = "{Machine-Learning Quantum States in the NISQ Era}",
    doi = "10.1146/annurev-conmatphys-031119-050651",
    journal = "Ann. Rev. Condensed Matter Phys.",
    volume = "11",
    number = "1",
    pages = "325--344",
    year = "2020"
}

@misc{matsuura2026krylovbreaktimesinhomogeneous,
      title={Krylov Break Times from an Inhomogeneous Lieb--Robinson Light Cone}, 
      author={Shunji Matsuura and Yoji Kawamura and Joseph Salfi and Satoshi Iso},
      year={2026},
      eprint={2608.27399},
      archivePrefix={arXiv},
      primaryClass={quant-ph},
      url={https://arxiv.org/abs/2608.27399}, 
}

@inproceedings{chen2018neuralode,
author = {Chen, Ricky T. Q. and Rubanova, Yulia and Bettencourt, Jesse and Duvenaud, David},
title = {Neural ordinary differential equations},
year = {2018},
publisher = {Curran Associates Inc.},
address = {Red Hook, NY, USA},
booktitle = {Proceedings of the 32nd International Conference on Neural Information Processing Systems},
pages = {6572–6583},
numpages = {12},
location = {Montr\'{e}al, Canada},
series = {NIPS'18},
url = {https://dl.acm.org/doi/10.5555/3327757.3327764}
}

@article{yoshioka2024hunting,
  title={Hunting for quantum-classical crossover in condensed matter problems},
  author={Yoshioka, Nobuyuki and Okubo, Tsuyoshi and Suzuki, Yasunari and Koizumi, Yuki and Mizukami, Wataru},
  journal={npj Quantum Information},
  volume={10},
  number={1},
  pages={45},
  year={2024},
  publisher={Nature Publishing Group UK London}
}

@article{z126-zdqj,
  title = {Computing $n$-Time Correlation Functions without Ancilla Qubits},
  author = {Wang, Xiaoyang and Xiong, Long and Cai, Xiaoxia and Yuan, Xiao},
  journal = {Phys. Rev. Lett.},
  volume = {135},
  issue = {23},
  pages = {230602},
  numpages = {9},
  year = {2025},
  month = {Dec},
  publisher = {American Physical Society},
  doi = {10.1103/z126-zdqj},
  url = {https://link.aps.org/doi/10.1103/z126-zdqj}
}

@misc{shinjo2024unveilingcleantwodimensionaldiscrete,
      title={Unveiling clean two-dimensional discrete time quasicrystals on a digital quantum computer}, 
      author={Kazuya Shinjo and Kazuhiro Seki and Tomonori Shirakawa and Rong-Yang Sun and Seiji Yunoki},
      year={2024},
      eprint={2403.16718},
      archivePrefix={arXiv},
      primaryClass={quant-ph},
      url={https://arxiv.org/abs/2403.16718}, 
}

@article{JinZhaoSun2025,
	author = {Sun, Jinzhao and Vilchez-Estevez, Lucia and Vedral, Vlatko and Boothroyd, Andrew T. and Kim, M. S.},
	da = {2025/02/06},
	doi = {10.1038/s41467-025-55955-2},
	id = {Sun2025},
	isbn = {2041-1723},
	journal = {Nature Communications},
	number = {1},
	pages = {1403},
	title = {Probing spectral features of quantum many-body systems with quantum simulators},
	ty = {JOUR},
	url = {https://doi.org/10.1038/s41467-025-55955-2},
	volume = {16},
	year = {2025}}

@article{Neill:2021wla,
    author = "Neill, C. and others",
    title = "{Accurately computing the electronic properties of a quantum ring}",
    doi = "10.1038/s41586-021-03576-2",
    journal = "Nature",
    volume = "594",
    number = "7864",
    pages = "508--512",
    year = "2021"
}

@article{King:2025mgn,
    author = "King, Andrew D. and others",
    title = "{Beyond-classical computation in quantum simulation}",
    doi = "10.1126/science.ado6285",
    journal = "Science",
    volume = "388",
    number = "6743",
    pages = "ado6285",
    year = "2025"
}

@article{PhysRevLett.106.050405,
  title = {Strong and Weak Thermalization of Infinite Nonintegrable Quantum Systems},
  author = {Ba\~nuls, M. C. and Cirac, J. I. and Hastings, M. B.},
  journal = {Phys. Rev. Lett.},
  volume = {106},
  issue = {5},
  pages = {050405},
  numpages = {4},
  year = {2011},
  month = {Feb},
  publisher = {American Physical Society},
  doi = {10.1103/PhysRevLett.106.050405},
  url = {https://link.aps.org/doi/10.1103/PhysRevLett.106.050405}
}

@article{FUNAHASHI1989183,
	author = {Ken-Ichi Funahashi},
	doi = {https://doi.org/10.1016/0893-6080(89)90003-8},
	issn = {0893-6080},
	journal = {Neural Networks},
	number = {3},
	pages = {183-192},
	title = {On the approximate realization of continuous mappings by neural networks},
	url = {https://www.sciencedirect.com/science/article/pii/0893608089900038},
	volume = {2},
	year = {1989}}

@article{Cybenko1989,
	author = {Cybenko, G. },
	da = {1989/12/01},
	doi = {10.1007/BF02551274},
	id = {Cybenko1989},
	isbn = {1435-568X},
	journal = {Mathematics of Control, Signals and Systems},
	number = {4},
	pages = {303--314},
	title = {Approximation by superpositions of a sigmoidal function},
	ty = {JOUR},
	url = {https://doi.org/10.1007/BF02551274},
	volume = {2},
	year = {1989}}

@article{Daley_22,
	author = {Daley, Andrew J. and Bloch, Immanuel and Kokail, Christian and Flannigan, Stuart and Pearson, Natalie and Troyer, Matthias and Zoller, Peter},
	da = {2022/07/01},
	doi = {10.1038/s41586-022-04940-6},
	id = {Daley2022},
	isbn = {1476-4687},
	journal = {Nature},
	number = {7920},
	pages = {667--676},
	title = {Practical quantum advantage in quantum simulation},
	ty = {JOUR},
	url = {https://doi.org/10.1038/s41586-022-04940-6},
	volume = {607},
	year = {2022}}

@article{Ghim:2024pxe,
    author = "Ghim, Dongwook and Honda, Masazumi",
    title = "{Digital Quantum Simulation for Spectroscopy of Schwinger Model}",
    eprint = "2404.14788",
    archivePrefix = "arXiv",
    primaryClass = "hep-lat",
    reportNumber = "RIKEN-iTHEMS-Report-24",
    doi = "10.22323/1.453.0213",
    journal = "PoS",
    volume = "LATTICE2023",
    pages = "213",
    year = "2024"
}

@article{PhysRevX.11.041039,
  title = {Randomized Compiling for Scalable Quantum Computing on a Noisy Superconducting Quantum Processor},
  author = {Hashim, Akel and Naik, Ravi K. and Morvan, Alexis and Ville, Jean-Loup and Mitchell, Bradley and Kreikebaum, John Mark and Davis, Marc and Smith, Ethan and Iancu, Costin and O'Brien, Kevin P. and Hincks, Ian and Wallman, Joel J. and Emerson, Joseph and Siddiqi, Irfan},
  journal = {Phys. Rev. X},
  volume = {11},
  issue = {4},
  pages = {041039},
  numpages = {12},
  year = {2021},
  month = {Nov},
  publisher = {American Physical Society},
  doi = {10.1103/PhysRevX.11.041039},
  url = {https://link.aps.org/doi/10.1103/PhysRevX.11.041039}
}

@article{Ezzel2023,
  title = {Dynamical decoupling for superconducting qubits: A performance survey},
  author = {Ezzell, N. and Pokharel, B. and Tewala, L. and Quiroz, G. and Lidar, D. A.},
  journal = {Phys. Rev. Appl.},
  volume = {20},
  issue = {6},
  pages = {064027},
  numpages = {42},
  year = {2023},
  month = {Dec},
  publisher = {American Physical Society},
  doi = {10.1103/PhysRevApplied.20.064027},
  url = {https://link.aps.org/doi/10.1103/PhysRevApplied.20.064027}
}

@article{SUZUKI1990319,
	author = {Masuo Suzuki},
	doi = {https://doi.org/10.1016/0375-9601(90)90962-N},
	issn = {0375-9601},
	journal = {Physics Letters A},
	number = {6},
	pages = {319-323},
	title = {Fractal decomposition of exponential operators with applications to many-body theories and Monte Carlo simulations},
	url = {https://www.sciencedirect.com/science/article/pii/037596019090962N},
	volume = {146},
	year = {1990}}

@article{Lorenzo_2024,
  title = {Robust Measurements of $n$-Point Correlation Functions of Driven-Dissipative Quantum Systems on a Digital Quantum Computer},
  author = {Del Re, Lorenzo and Rost, Brian and Foss-Feig, Michael and Kemper, A. F. and Freericks, J. K.},
  journal = {Phys. Rev. Lett.},
  volume = {132},
  issue = {10},
  pages = {100601},
  numpages = {6},
  year = {2024},
  month = {Mar},
  publisher = {American Physical Society},
  doi = {10.1103/PhysRevLett.132.100601},
  url = {https://link.aps.org/doi/10.1103/PhysRevLett.132.100601}
}

@article{10.1145/3408039,
author = {Flammia, Steven T. and Wallman, Joel J.},
title = {Efficient Estimation of Pauli Channels},
year = {2020},
issue_date = {December 2020},
publisher = {Association for Computing Machinery},
address = {New York, NY, USA},
volume = {1},
number = {1},
url = {https://doi.org/10.1145/3408039},
doi = {10.1145/3408039},
journal = {ACM Transactions on Quantum Computing},
month = dec,
articleno = {3},
numpages = {32}
}

@misc{kingma2017adammethodstochasticoptimization,
      title={Adam: A Method for Stochastic Optimization}, 
      author={Diederik P. Kingma and Jimmy Ba},
      year={2017},
      eprint={1412.6980},
      archivePrefix={arXiv},
      primaryClass={cs.LG},
      url={https://arxiv.org/abs/1412.6980}, 
}

@book{john1982partial,
  title     = {Partial Differential Equations},
  author    = {John, Fritz},
  series    = {Applied Mathematical Sciences},
  volume    = {1},
  edition   = {4th},
  year      = {1982},
  publisher = {Springer-Verlag},
  address   = {New York},
  isbn      = {0-387-90609-6},
  url       = {https://link.springer.com/book/9780387906096}
}

@article{ORUS2014117,
	author = {Rom{\'a}n Or{\'u}s},
	doi = {https://doi.org/10.1016/j.aop.2014.06.013},
	issn = {0003-4916},
	journal = {Annals of Physics},
	pages = {117-158},
	title = {A practical introduction to tensor networks: Matrix product states and projected entangled pair states},
	url = {https://www.sciencedirect.com/science/article/pii/S0003491614001596},
	volume = {349},
	year = {2014}}

@article{Berezutskii_25,
	author = {Berezutskii, Aleksandr and Liu, Minzhao and Acharya, Atithi and Ellerbrock, Roman and Gray, Johnnie and Haghshenas, Reza and He, Zichang and Khan, Abid and Kuzmin, Viacheslav and Lyakh, Dmitry and Lykov, Danylo and Mandr{\`a}, Salvatore and Mansell, Christopher and Melnikov, Alexey and Melnikov, Artem and Mironov, Vladimir and Morozov, Dmitry and Neukart, Florian and Nocera, Alberto and Perlin, Michael A. and Perelshtein, Michael and Steinberg, Matthew and Shaydulin, Ruslan and Villalonga, Benjamin and Pflitsch, Markus and Pistoia, Marco and Vinokur, Valerii and Alexeev, Yuri},
	da = {2025/10/01},
	doi = {10.1038/s42254-025-00853-1},
	id = {Berezutskii2025},
	isbn = {2522-5820},
	journal = {Nature Reviews Physics},
	number = {10},
	pages = {581--593},
	title = {Tensor networks for quantum computing},
	ty = {JOUR},
	url = {https://doi.org/10.1038/s42254-025-00853-1},
	volume = {7},
	year = {2025}}

@inbook{3454287.3454765,
author = {Rubanova, Yulia and Chen, Ricky T. Q. and Duvenaud, David},
title = {Latent ODEs for irregularly-sampled time series},
year = {2019},
publisher = {Curran Associates Inc.},
address = {Red Hook, NY, USA},
booktitle = {Proceedings of the 33rd International Conference on Neural Information Processing Systems},
articleno = {478},
numpages = {11}
}

@book{gubernatis2016quantum,
  title={Quantum Monte Carlo Methods: Algorithms for Lattice Models},
  author={Gubernatis, J. and Kawashima, N. and Werner, P.},
  isbn={9781316483121},
  url={https://books.google.co.jp/books?id=KKMtDAAAQBAJ},
  year={2016},
  publisher={Cambridge University Press}
}

@article{jt8s-hzhd,
  title = {Quantum computing quantum Monte Carlo algorithm},
  author = {Zhang, Yukun and Huang, Yifei and Sun, Jinzhao and Lv, Dingshun and Yuan, Xiao},
  journal = {Phys. Rev. A},
  volume = {112},
  issue = {2},
  pages = {022428},
  numpages = {28},
  year = {2025},
  month = {Aug},
  publisher = {American Physical Society},
  doi = {10.1103/jt8s-hzhd},
  url = {https://link.aps.org/doi/10.1103/jt8s-hzhd}
}

@article{Huggins2022,
	author = {Huggins, William J. and O'Gorman, Bryan A. and Rubin, Nicholas C. and Reichman, David R. and Babbush, Ryan and Lee, Joonho},
	da = {2022/03/01},
	doi = {10.1038/s41586-021-04351-z},
	id = {Huggins2022},
	isbn = {1476-4687},
	journal = {Nature},
	number = {7901},
	pages = {416--420},
	title = {Unbiasing fermionic quantum Monte Carlo with a quantum computer},
	ty = {JOUR},
	url = {https://doi.org/10.1038/s41586-021-04351-z},
	volume = {603},
	year = {2022}}

@article{Abanin2025,
	author = {Abanin, Dmitry A. and Acharya, Rajeev and Aghababaie-Beni, Laleh and Aigeldinger, Georg and Ajoy, Ashok and Alcaraz, Ross and Aleiner, Igor and Andersen, Trond I. and Ansmann, Markus and Arute, Frank and Arya, Kunal and Asfaw, Abraham and Astrakhantsev, Nikita and Atalaya, Juan and Babbush, Ryan and Bacon, Dave and Ballard, Brian and Bardin, Joseph C. and Bengs, Christian and Bengtsson, Andreas and Bilmes, Alexander and Boixo, Sergio and Bortoli, Gina and Bourassa, Alexandre and Bovaird, Jenna and Bowers, Dylan and Brill, Leon and Broughton, Michael and Browne, David A. and Buchea, Brett and Buckley, Bob B. and Buell, David A. and Burger, Tim and Burkett, Brian and Bushnell, Nicholas and Cabrera, Anthony and Campero, Juan and Chang, Hung-Shen and Chen, Yu and Chen, Zijun and Chiaro, Ben and Chih, Liang-Ying and Chik, Desmond and Chou, Charina and Claes, Jahan and Cleland, Agnetta Y. and Cogan, Josh and Cohen, Saul and Collins, Roberto and Conner, Paul and Courtney, William and Crook, Alexander L. and Curtin, Ben and Das, Sayan and De Lorenzo, Laura and Debroy, Dripto M. and Demura, Sean and Devoret, Michel and Di Paolo, Agustin and Donohoe, Paul and Drozdov, Ilya and Dunsworth, Andrew and Earle, Clint and Eickbusch, Alec and Elbag, Aviv Moshe and Elzouka, Mahmoud and Erickson, Catherine and Faoro, Lara and Farhi, Edward and Ferreira, Vinicius S. and Burgos, Leslie Flores and Forati, Ebrahim and Fowler, Austin G. and Foxen, Brooks and Ganjam, Suhas and Garcia, Gonzalo and Gasca, Robert and Genois, {\'E}lie and Giang, William and Gidney, Craig and Gilboa, Dar and Gosula, Raja and Dau, Alejandro Grajales and Graumann, Dietrich and Greene, Alex and Gross, Jonathan A. and Gu, Hanfeng and Habegger, Steve and Hall, John and Hamamura, Ikko and Hamilton, Michael C. and Hansen, Monica and Harrigan, Matthew P. and Harrington, Sean D. and Heslin, Stephen and Heu, Paula and Higgott, Oscar and Hill, Gordon and Hilton, Jeremy and Hong, Sabrina and Huang, Hsin-Yuan and Huff, Ashley and Huggins, William J. and Ioffe, Lev B. and Isakov, Sergei V. and Iveland, Justin and Jeffrey, Evan and Jiang, Zhang and Jin, Xiaoxuan and Jones, Cody and Jordan, Stephen and Joshi, Chaitali and Juhas, Pavol and Kabel, Andreas and Kafri, Dvir and Kang, Hui and Karamlou, Amir H. and Kechedzhi, Kostyantyn and Kelly, Julian and Khaire, Trupti and Khattar, Tanuj and Khezri, Mostafa and Kim, Seon and King, Robbie and Klimov, Paul V. and Klots, Andrey R. and Kobrin, Bryce and Korotkov, Alexander N. and Kostritsa, Fedor and Kothari, Robin and Kreikebaum, John Mark and Kurilovich, Vladislav D. and Kyoseva, Elica and Landhuis, David and Lange-Dei, Tiano and Langley, Brandon W. and Laptev, Pavel and Lau, Kim-Ming and Le Guevel, Lo{\"\i}ck and Ledford, Justin and Lee, Joonho and Lee, Kenny and Lensky, Yuri D. and Leon, Shannon and Lester, Brian J. and Li, Wing Yan and Lill, Alexander T. and Liu, Wayne and Livingston, William P. and Locharla, Aditya and Lucero, Erik and Lundahl, Daniel and Lunt, Aaron and Madhuk, Sid and Malone, Fionn D. and Maloney, Ashley and Mandr{\`a}, Salvatore and Manyika, James M. and Martin, Leigh S. and Martin, Orion and Martin, Steven and Matias, Yossi and Maxfield, Cameron and McClean, Jarrod R. and McEwen, Matt and Meeks, Seneca and Megrant, Anthony and Mi, Xiao and Miao, Kevin C. and Mieszala, Amanda and Minev, Zlatko and Molavi, Reza and Molina, Sebastian and Montazeri, Shirin and Morvan, Alexis and Movassagh, Ramis and Mruczkiewicz, Wojciech and Naaman, Ofer and Neeley, Matthew and Neill, Charles and Nersisyan, Ani and Neven, Hartmut and Newman, Michael and Ng, Jiun How and Nguyen, Anthony and Nguyen, Murray and Ni, Chia-Hung and Niu, Murphy Yuezhen and Oas, Logan and O'Brien, Thomas E. and Oliver, William D. and Opremcak, Alex and Ottosson, Kristoffer and Petukhov, Andre and Pizzuto, Alex and Platt, John and Potter, Rebecca and Pritchard, Orion and Pryadko, Leonid P. and Quintana, Chris and Ramachandran, Ganesh and Ramanathan, Chandrasekhar and Reagor, Matthew J. and Redding, John and Rhodes, David M. and Roberts, Gabrielle and Rosenberg, Eliott and Rosenfeld, Emma and Roushan, Pedram and Rubin, Nicholas C. and Saei, Negar and Sank, Daniel and Sankaragomathi, Kannan and Satzinger, Kevin J. and Schmidhuber, Alexander and Schurkus, Henry F. and Schuster, Christopher and Schuster, Thomas and Shearn, Michael J. and Shorter, Aaron and Shutty, Noah and Shvarts, Vladimir and Sivak, Volodymyr and Skruzny, Jindra and Small, Spencer and Smelyanskiy, Vadim and Smith, W. Clarke and Somma, Rolando D. and Springer, Sofia and Sterling, George and Strain, Doug and Suchard, Jordan and Suchsland, Philippe and Szasz, Aaron and Sztein, Alex and Thor, Douglas and Tomita, Eifu and Torres, Alfredo and Torunbalci, M. Mert and Vaishnav, Abeer and Vargas, Justin and Vdovichev, Sergey and Vidal, Guifre and Villalonga, Benjamin and Heidweiller, Catherine Vollgraff and Waltman, Steven and Wang, Shannon X. and Ware, Brayden and Weber, Kate and Weidel, Travis and Westerhout, Tom and White, Theodore and Wong, Kristi and Woo, Bryan W. K. and Xing, Cheng and Yao, Z. Jamie and Yeh, Ping and Ying, Bicheng and Yoo, Juhwan and Yosri, Noureldin and Young, Grayson and Zalcman, Adam and Zhang, Chongwei and Zhang, Yaxing and Zhu, Ningfeng and Zobrist, Nicholas and Google Quantum AI and Collaborators},
	da = {2025/10/01},
	doi = {10.1038/s41586-025-09526-6},
	id = {Abanin2025},
	isbn = {1476-4687},
	journal = {Nature},
	number = {8086},
	pages = {825--830},
	title = {Observation of constructive interference at the edge of quantum ergodicity},
	ty = {JOUR},
	url = {https://doi.org/10.1038/s41586-025-09526-6},
	volume = {646},
	year = {2025}}

@misc{liang2024observationanomalousinformationscrambling,
      title={Observation of anomalous information scrambling in a Rydberg atom array}, 
      author={Xinhui Liang and Zongpei Yue and Yu-Xin Chao and Zhen-Xing Hua and Yige Lin and Meng Khoon Tey and Li You},
      year={2024},
      eprint={2410.16174},
      archivePrefix={arXiv},
      primaryClass={quant-ph},
      url={https://arxiv.org/abs/2410.16174}, 
}

@article{PhysRevLett.129.160602,
  title = {Probing Operator Spreading via Floquet Engineering in a Superconducting Circuit},
  author = {Zhao, S. K. and Ge, Zi-Yong and Xiang, Zhongcheng and Xue, G. M. and Yan, H. S. and Wang, Z. T. and Wang, Zhan and Xu, H. K. and Su, F. F. and Yang, Z. H. and Zhang, He and Zhang, Yu-Ran and Guo, Xue-Yi and Xu, Kai and Tian, Ye and Yu, H. F. and Zheng, D. N. and Fan, Heng and Zhao, S. P.},
  journal = {Phys. Rev. Lett.},
  volume = {129},
  issue = {16},
  pages = {160602},
  numpages = {7},
  year = {2022},
  month = {Oct},
  publisher = {American Physical Society},
  doi = {10.1103/PhysRevLett.129.160602},
  url = {https://link.aps.org/doi/10.1103/PhysRevLett.129.160602}
}

@article{Swingle2018,
	author = {Swingle, Brian},
	da = {2018/10/01},
	doi = {10.1038/s41567-018-0295-5},
	id = {Swingle2018},
	isbn = {1745-2481},
	journal = {Nature Physics},
	number = {10},
	pages = {988--990},
	title = {Unscrambling the physics of out-of-time-order correlators},
	ty = {JOUR},
	url = {https://doi.org/10.1038/s41567-018-0295-5},
	volume = {14},
	year = {2018}}

@article{PhysRevA.105.032420,
  title = {Algebraic compression of quantum circuits for Hamiltonian evolution},
  author = {K\"okc\"u, Efekan and Camps, Daan and Bassman Oftelie, Lindsay and Freericks, J. K. and de Jong, Wibe A. and Van Beeumen, Roel and Kemper, Alexander F.},
  journal = {Phys. Rev. A},
  volume = {105},
  issue = {3},
  pages = {032420},
  numpages = {14},
  year = {2022},
  month = {Mar},
  publisher = {American Physical Society},
  doi = {10.1103/PhysRevA.105.032420},
  url = {https://link.aps.org/doi/10.1103/PhysRevA.105.032420}
}

@misc{Qiskit,
      title={Quantum computing with Qiskit}, 
      author={Ali Javadi-Abhari and Matthew Treinish and Kevin Krsulich and Christopher J. Wood and Jake Lishman and Julien Gacon and Simon Martiel and Paul D. Nation and Lev S. Bishop and Andrew W. Cross and Blake R. Johnson and Jay M. Gambetta},
      year={2024},
      eprint={2405.08810},
      archivePrefix={arXiv},
      primaryClass={quant-ph},
      url={https://arxiv.org/abs/2405.08810}, 
}

@article{Lloyd:1996aai,
    author = "Lloyd, Seth",
    title = "{Universal Quantum Simulators}",
    doi = "10.1126/science.273.5278.1073",
    journal = "Science",
    volume = "273",
    number = "5278",
    pages = "1073",
    year = "1996"
}

@article{Kempe:2004sak,
    author = "Kempe, Julia and Kitaev, Alexei and Regev, Oded",
    title = "{The Complexity of the Local Hamiltonian Problem}",
    eprint = "quant-ph/0406180",
    archivePrefix = "arXiv",
    doi = "10.1137/S0097539704445226",
    journal = "SIAM J. Comput.",
    volume = "35",
    number = "5",
    pages = "1070--1097",
    year = "2006"
}

@article{PhysRevA.65.042323,
  title = {Simulating physical phenomena by quantum networks},
  author = {Somma, R. and Ortiz, G. and Gubernatis, J. E. and Knill, E. and Laflamme, R.},
  journal = {Phys. Rev. A},
  volume = {65},
  issue = {4},
  pages = {042323},
  numpages = {17},
  year = {2002},
  month = {Apr},
  publisher = {American Physical Society},
  doi = {10.1103/PhysRevA.65.042323},
  url = {https://link.aps.org/doi/10.1103/PhysRevA.65.042323}
}

@article{PhysRevLett.113.020505,
  title = {Efficient Quantum Algorithm for Computing $n$-time Correlation Functions},
  author = {Pedernales, J. S. and Di Candia, R. and Egusquiza, I. L. and Casanova, J. and Solano, E.},
  journal = {Phys. Rev. Lett.},
  volume = {113},
  issue = {2},
  pages = {020505},
  numpages = {5},
  year = {2014},
  month = {Jul},
  publisher = {American Physical Society},
  doi = {10.1103/PhysRevLett.113.020505},
  url = {https://link.aps.org/doi/10.1103/PhysRevLett.113.020505}
}

@article{PhysRevD.104.074518,
  title = {Diagnosis of information scrambling from Hamiltonian evolution under decoherence},
  author = {Hayata, Tomoya and Hidaka, Yoshimasa and Kikuchi, Yuta},
  journal = {Phys. Rev. D},
  volume = {104},
  issue = {7},
  pages = {074518},
  numpages = {25},
  year = {2021},
  month = {Oct},
  publisher = {American Physical Society},
  doi = {10.1103/PhysRevD.104.074518},
  url = {https://link.aps.org/doi/10.1103/PhysRevD.104.074518}
}

@article{PRXQuantum.3.040309,
  title = {Dynamical Hadron Formation in Long-Range Interacting Quantum Spin Chains},
  author = {Vovrosh, Joseph and Mukherjee, Rick and Bastianello, Alvise and Knolle, Johannes},
  journal = {PRX Quantum},
  volume = {3},
  issue = {4},
  pages = {040309},
  numpages = {12},
  year = {2022},
  month = {Oct},
  publisher = {American Physical Society},
  doi = {10.1103/PRXQuantum.3.040309},
  url = {https://link.aps.org/doi/10.1103/PRXQuantum.3.040309}
}

@article{PhysRevLett.122.150601,
  title = {Confined Quasiparticle Dynamics in Long-Range Interacting Quantum Spin Chains},
  author = {Liu, Fangli and Lundgren, Rex and Titum, Paraj and Pagano, Guido and Zhang, Jiehang and Monroe, Christopher and Gorshkov, Alexey V.},
  journal = {Phys. Rev. Lett.},
  volume = {122},
  issue = {15},
  pages = {150601},
  numpages = {7},
  year = {2019},
  month = {Apr},
  publisher = {American Physical Society},
  doi = {10.1103/PhysRevLett.122.150601},
  url = {https://link.aps.org/doi/10.1103/PhysRevLett.122.150601}
}

@misc{zhai2025universalquantumcomputationalspectroscopy,
      title={Universal Quantum Computational Spectroscopy on a Quantum Chip}, 
      author={Chonghao Zhai and Jinzhao Sun and Jieshan Huang and Jun Mao and Hongchang Bao and Siyuan Zhang and Qihuang Gong and Vlatko Vedral and Xiao Yuan and Jianwei Wang},
      year={2025},
      eprint={2506.22418},
      archivePrefix={arXiv},
      primaryClass={quant-ph},
      url={https://arxiv.org/abs/2506.22418}, 
}

@article{Efekan_2024,
	author = {K{\"o}kc{\"u}, Efekan and Labib, Heba A. and Freericks, J. K. and Kemper, A. F.},
	da = {2024/05/08},
	doi = {10.1038/s41467-024-47729-z},
	id = {K{\"o}kc{\"u}2024},
	isbn = {2041-1723},
	journal = {Nature Communications},
	number = {1},
	pages = {3881},
	title = {A linear response framework for quantum simulation of bosonic and fermionic correlation functions},
	ty = {JOUR},
	url = {https://doi.org/10.1038/s41467-024-47729-z},
	volume = {15},
	year = {2024}}

@book{bishop2023deep,
  title={Deep learning: Foundations and concepts},
  author={Bishop, Christopher M and Bishop, Hugh},
  year={2023},
  publisher={Springer Nature}
}

@article{Roggero_2019,
  title = {Dynamic linear response quantum algorithm},
  author = {Roggero, Alessandro and Carlson, Joseph},
  journal = {Phys. Rev. C},
  volume = {100},
  issue = {3},
  pages = {034610},
  numpages = {6},
  year = {2019},
  month = {Sep},
  publisher = {American Physical Society},
  doi = {10.1103/PhysRevC.100.034610},
  url = {https://link.aps.org/doi/10.1103/PhysRevC.100.034610}
}

@article{Kosugi_2020,
  title = {Construction of Green's functions on a quantum computer: Quasiparticle spectra of molecules},
  author = {Kosugi, Taichi and Matsushita, Yu-ichiro},
  journal = {Phys. Rev. A},
  volume = {101},
  issue = {1},
  pages = {012330},
  numpages = {12},
  year = {2020},
  month = {Jan},
  publisher = {American Physical Society},
  doi = {10.1103/PhysRevA.101.012330},
  url = {https://link.aps.org/doi/10.1103/PhysRevA.101.012330}
}

@article{Chen_2021,
  title = {Variational quantum eigensolver for dynamic correlation functions},
  author = {Chen, Hongxiang and Nusspickel, Max and Tilly, Jules and Booth, George H.},
  journal = {Phys. Rev. A},
  volume = {104},
  issue = {3},
  pages = {032405},
  numpages = {12},
  year = {2021},
  month = {Sep},
  publisher = {American Physical Society},
  doi = {10.1103/PhysRevA.104.032405},
  url = {https://link.aps.org/doi/10.1103/PhysRevA.104.032405}
}

@article{Ciavarella_2020,
  title = {Algorithm for quantum computation of particle decays},
  author = {Ciavarella, Anthony},
  journal = {Phys. Rev. D},
  volume = {102},
  issue = {9},
  pages = {094505},
  numpages = {24},
  year = {2020},
  month = {Nov},
  publisher = {American Physical Society},
  doi = {10.1103/PhysRevD.102.094505},
  url = {https://link.aps.org/doi/10.1103/PhysRevD.102.094505}
}

@inproceedings{Dinh:2016pgf,
    author = "Dinh, Laurent and Sohl-Dickstein, Jascha and Bengio, Samy",
    title = "{Density estimation using Real NVP}",
    booktitle={International Conference on Learning Representations},
    eprint = "1605.08803",
    archivePrefix = "arXiv",
    primaryClass = "cs.LG",
    year = "2017"
}
\bibliographystyle{apsrev4-2}

\section*{End Note}

\noindent \emph{Initial state preparation}~---~In this note, we present the quantum circuits to prepare the initial state $\rho_i=(\sigma_i+\mathbb{I})/d$. By preparing $\rho_i$, the Pauli coefficient $c_i(t)$ as the short-time training data can be measured, since the expectation value of $O$ in the time-evolved initial state is
\begin{align}
     \mathrm{Tr}\!\left[O\,e^{-iHt}\rho_ie^{iHt}\right]= \mathrm{Tr}\!\left[O(t)\,\rho_i\right]=c_i(t)+\frac{1}{d}\mathrm{Tr}[O],
     \label{eq:c_it_measurement}
\end{align}
where the last term $\mathrm{Tr}[O]$ vanishes if $O$ is written as a linear combination of non-identity Pauli strings. 

To prepare $\rho_i$, we firstly prepare a maximally mixed state $\mathbb{I}/d$ of the target quantum system. This can be achieved on quantum computers by either preparing the purified Bell state $\ket{\Phi^+}^{\otimes N}$ with $\ket{\Phi^+}:=(\ket{00}+\ket{11})/\sqrt{2}$, or sampling Pauli gates $\sigma_i\in\bos{P}_N$ with equal probability and applying them to an arbitrary initial state for every measurement repetition~\cite{White2009}. Then the state $\rho_i$ can be prepared by introducing an ancillary qubit and using the following quantum circuit:
\begin{align}
    \Qcircuit @C=1.2em @R=1.2em {
\lstick{\ket{0}_a} & \gate{H} & \ctrl{1} & \gate{H} & \meter &\\
\lstick{\mathbb{I}/d} & \qw & \gate{\sigma_i} & \qw &\qw& \Longrightarrow \rho_i.
}\nonumber
\end{align}
Here, the measurement on the ancillary qubit is in the Pauli-$Z$ basis. It can be shown that if the ancillary qubit after measurement is in the state $\ket{0}_a$, the output state of the system is $\rho_i$, and the success probability of measuring $\ket{0}$ is $1/2$, which has no dependence on the system size. Therefore, the initial state $\rho_i$ can be prepared efficiently on quantum computers.

For the hardware implementation in the main text, we prepare the Bell state on each system qubit of the $N$-qubit Ising spin chain, which requires additional $N+1$ ancillary qubits. Therefore, a total of $2N+1$ qubits are used in the hardware implementation.

\vspace{1.5mm}
\noindent \emph{Evolution equation of local Pauli coefficients}~---~We derive the evolution equation of local Pauli coefficients $\mathbf{h}(t)$ in Eq.~\eqref{eq:truncated-ODE-2} of the main text, which guides the design of the FAN used in the operator learning method. Taking the operator decomposition in Eq.~\eqref{eq:Ot-expansion} into the Heisenberg equation 
\begin{align}
    \frac{dO(t)}{dt}=-i[O(t),H],
\end{align}
and using the orthogonality relation $\Tr(\sigma_i\sigma_j)/d=\delta_{ij}$, we derive the evolution equation of full Pauli coefficients
\begin{align}
    \frac{dc_i(t)}{dt}=\sum_{\sigma_j\in \PN}L_{ij}\;c_j(t),
    \label{eq:c_i-full-evolution}
\end{align}
where $L_{ij}:=-i\Tr(H[\sigma_i,\sigma_j])/d$ is an anti-symmetric real matrix of dimension $4^N\times 4^N$. The local Pauli truncation retains only spatially local Pauli strings in $\PN$, such that the whole space of $\PN$ is divided into two parts
\begin{equation}
    \begin{aligned}
    P&=\{\text{retained local Pauli strings}\}; \\
    Q&=\{\text{truncated non-local Pauli strings}\},
\end{aligned}
\end{equation}
and the $4^N$ Pauli coefficients are divided into $(c_1(t),\ldots ,c_{4^N}(t))=(\mathbf{h}_P(t), \mathbf{h}_Q(t))^T$. Therefore, the matrix $L$ and Eq.~\eqref{eq:c_i-full-evolution} are partitioned into blocks as
\begin{align}
    \frac{d}{dt}\left( \begin{array}{c}
\mathbf{h}_P \\
\mathbf{h}_Q 
\end{array} \right)=\left( \begin{array}{cc}
L_{PP} & L_{PQ} \\
L_{QP} & L_{QQ} 
\end{array} \right)\left( \begin{array}{c}
\mathbf{h}_P \\
\mathbf{h}_Q 
\end{array} \right),
\end{align}
where both $L_{PP}$ and $L_{QQ} $ are anti-symmetric real matrix. Here, $L_{PP}$ is the Hamiltonian $H^{\tr}$ in the local Pauli basis, and $\mathbf{h}_P$ is the $\Ntr$-dimensional local coefficients $\mathbf{h}$ denoted in the main text.

Eliminating the non-local Pauli strings in $Q$ produces a memory kernel equation according to Duhamel's principle~\cite{john1982partial}
\begin{align}
    \frac{d\mathbf{h}_P(t)}{dt } = L_{PP}\mathbf{h}_P(t)+\int_{0}^{t}K(t-s)\mathbf{h}_P(s)ds,
    \label{eq:hp(t)}
\end{align}
where the kernel function $K(t-s)$ is an $\Ntr\times \Ntr$ matrix defined by
\begin{align}
    K(t-s):=L_{PQ}e^{L_{QQ}(t-s)} L_{QP}.
    \label{eq:K-expression}
\end{align}
Then, since Eq.~\eqref{eq:hp(t)} is a linear homogeneous differential equation, there exists a matrix $U(t)$ such that 
\begin{align}
    \mathbf{h}_P(t)= U(t)\mathbf{h}_P(0),
    \label{eq:hP-U-hP}
\end{align}
and $U(t)$ satisfies the differential equation 
\begin{align}
    \frac{d U(t)}{dt } = L_{PP}U(t)+\int_{0}^{t}K(t-s)U(s)ds
\end{align}
with an initial condition $U(0)=I$. Finally, taking Eq.~\eqref{eq:hP-U-hP} into the second term of Eq.~\eqref{eq:hp(t)} gives the evolution equation of local Pauli coefficients in Eq.~\eqref{eq:truncated-ODE-2} of the main text. The formal expression of the propagator matrix $G(t)$ is given by
\begin{align}
    G(t) = \int_{0}^{t}K(t-s) U(s)ds.
\end{align}
We see that $G(t)$ is hard to solve explicitly due to the memory kernel $K(t-s)$ in Eq.~\eqref{eq:K-expression} with an exponentially large $L_{QQ}$.

The memory kernel $K(t-s)$ and therefore the propagator matrix $G(t)$ contain many periodic modes. This can be seen by diagonalizing the anti-symmetric real matrix $L_{QQ} = \sum_a \bos{v}_a^{\dagger}  (i\omega_a) \bos{v}_a$, which gives the real eigenvalues $\omega_a$ and the corresponding eigenvector $\bos{v}_a$. We see that the kernel function oscillates as 
\begin{align}
    K(t-s)=\sum_{a} e^{i\omega_a(t-s)} L_{PQ}\bos{v}_a^{\dagger}\bos{v}_a L_{QP}.
    \label{eq:K-decomposition}
\end{align}
In the operator learning method, we use FAN to approximate these periodic modes in $K(t-s)$. The approximation is accurate if only a few $\omega_a$ contribute to Eq.~\eqref{eq:K-decomposition}, or if the coupling between the local and non-local Pauli strings $L_{QP/PQ}$ is small. These two conditions are satisfied if the operator dynamics $O(t)$ have a sparse spectrum, or if the Hamiltonian has small spatial coupling strength, such that the leakage from space $P$ to $Q$ by $L_{PQ}$ is negligible. On the other hand, the FAN has large prediction error if the operator dynamics have dense spectrum, or if a many-body system has strong spatial couplings. These limitations of FAN can be mitigated by increasing the training data size, as numerically studied in Supplemental Material~\cite{supp}.

\nocite{PhysRevLett.97.050401,matsuura2026krylovbreaktimesinhomogeneous,10.1145/3408039,Ezzel2023,PhysRevX.11.041039,kingma2017adammethodstochasticoptimization,3454287.3454765}
\clearpage
\newpage

\setcounter{page}{1}
\renewcommand{\thefigure}{S\arabic{figure}}
\renewcommand{\theequation}{S\arabic{equation}}



\appendix

\section{Operator truncation and symmetry reduction} \label{app:trunc}
For an $N$-qubit quantum many-body system, its time-evolved operator can be expanded in the complete Pauli basis $\PN$ as
\begin{align}
    O(t) = \sum_{\sigma_i\in\PN} c_i(t) \,\sigma_i,
    \label{eq:Ot-expansion}
\end{align}
where $c_i(t)$ is the Pauli coefficient measured using quantum devices. However, measuring the complete Pauli basis $\PN$ requires an exponential time complexity of $\OO(4^N)$, and the exponentially large number of coefficients $c_i(t)$ as training data also makes it impractical to input to the neural network. In the following content, we introduce a \textit{local Pauli truncation} to reduce the exponential time complexity in the short-time data acquisition and long-time prediction processes. Additionally, targeting the specific quantum system, the Pauli coefficients can be further reduced by the system symmetry.

\subsection{Local Pauli truncation in short-time data}
\label{Sec3:truncation}
In the main text, we assume that the quantum many-body system is evolved by a spatially $k$-local Hamiltonian, and the observable $O$ can also be written as a linear combination of local operators. Additionally, we only need to measure the short-time behavior of the coefficient $c_i(t)$. These assumptions allow us to perform the local truncation of $\sigma_i$ in Eq.~\eqref{eq:Ot-expansion} during the data acquisition process as guaranteed by the Lieb-Robinson theorem~\cite{PhysRevLett.97.050401,matsuura2026krylovbreaktimesinhomogeneous}.

For two operators $O_A$ and $O_B$ supported on local regions $A$ and $B$ respectively, the Lieb-Robinson theorem states that their commutator at time $t$ satisfies
\begin{align}
\|[O_A(t),O_B]\|
\le
C\,\exp\!\left[-\mu\big(l(A,B)-v|t|\big)\right],
\label{eq:lieb-robinson-theorem}
\end{align}
where $l(A, B)$ denotes the distance between the two regions, $v$ is the Lieb-Robinson velocity, and $C,\mu$ are positive constants depending on the interaction strength and range.

This result implies that the support of an operator evolved in the Heisenberg picture spreads only within an effective light cone whose slope is determined by $v$. In the expansion of the time-evolved $O(t)$ in Eq.~\eqref{eq:Ot-expansion}, the coefficients of $\sigma_i$s whose supports lie outside the light cone are exponentially suppressed. Specifically, if a Pauli string $\sigma_i$ acts on sites separated by a distance $l$ to the local $O$, its coefficient is exponentially suppressed by $l$ as
\begin{align}
|c_i(t)| \lesssim e^{-\mu(l-vt)} .
\end{align}
Therefore, for a finite evolution time $t$, the dominant contributions to the operator expansion come from Pauli strings supported in the vicinity of the local $O$ with the radius $r\sim vt$. For the short-time evolution considered in the learning task, the operator dynamics can be accurately approximated by Pauli strings acting only on local sites. This local truncation of Pauli strings reduces the quantum and classical time complexity from exponential to polynomial.

As an example, consider a quantum system on a $D$-dimensional lattice, and $O$ is a linear combination of $k$-local observables across the whole quantum system. Given a short evolution time up to $T$, the number of Pauli bases after the local truncation reads
\begin{align}
    N_O^{\mathrm{tr}} = N\times 4^{k(2vT)^D},
    \label{eq:Ntr-scaling}
\end{align}
where $4^{k(2vT)^D}$ is the total number of Pauli bases within the range $vT$ of each $k$-local observables. This number increases linearly with the system size $N$, ensuring the scalability of the data acquisition process of the operator learning method.

\begin{figure*}
    \centering
    \includegraphics[width=0.95\textwidth]
    {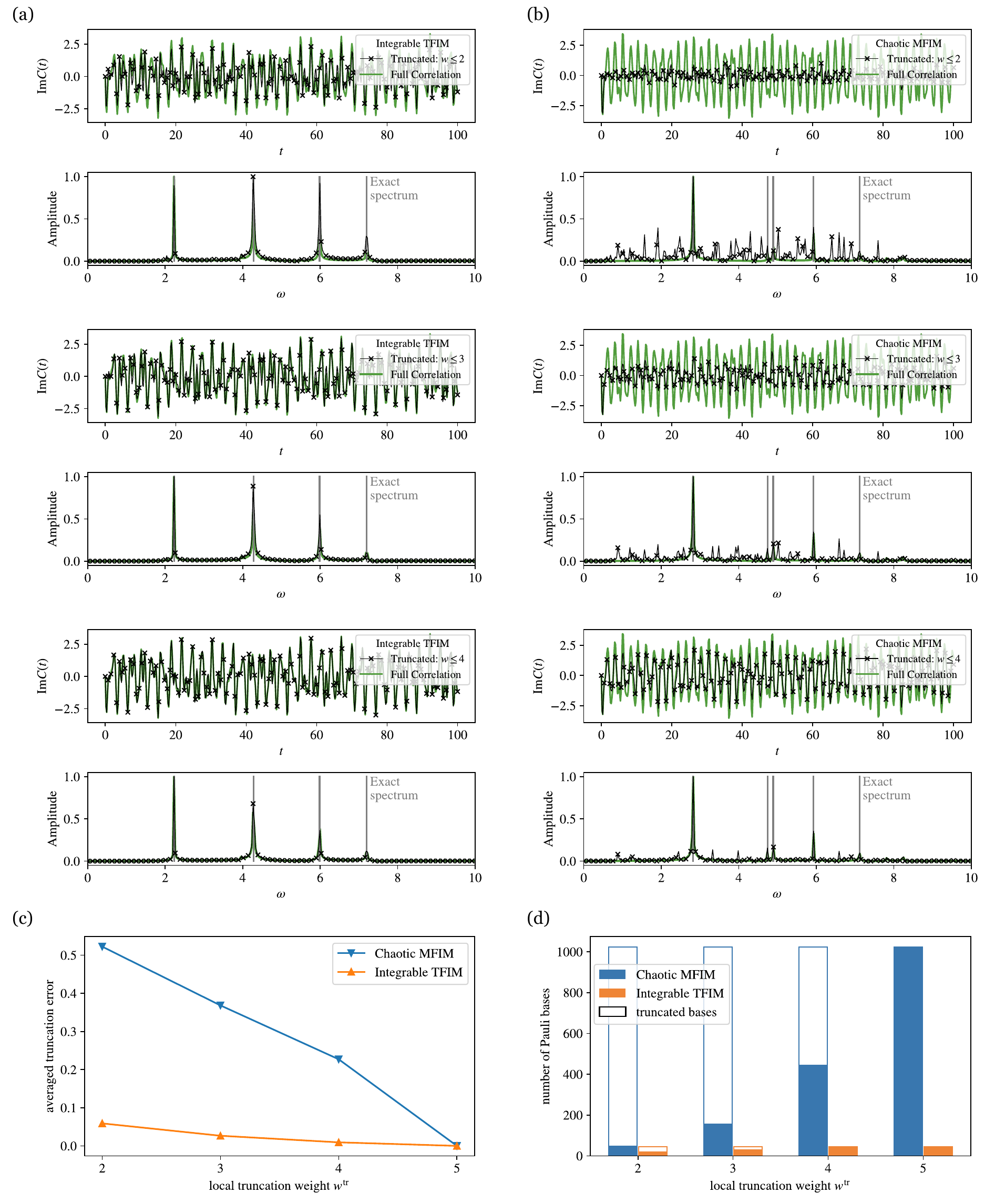}
    \caption{
    Effect of the local Pauli truncation on the correlation function $C(t)$ of the (a) integrable transverse-field Ising model (TFIM) and (b) chaotic mixed-field Ising model (MFIM) with $N=5$ spins. $w$ is the weight of local Pauli bases retained to reconstruct $C_{\tr}(t)$. (c) The averaged truncation error $\overline{|[C_{\tr:w\le w^{\tr}}(t)-C(t)]/C(t)|}$ over time $t\in[0,100]$ versus the local truncation weight $w^{\tr}$. $w^{\tr}=5$ denotes the case without local truncation in the $N=5$ spin system. (d) The number of Pauli bases retained after the local Pauli truncation for different truncation weights $w^{\tr}$ in MFIM and TFIM.
    }
    \label{Fig:LPT-effect}
\end{figure*}

\subsection{Local Pauli truncation in long-time prediction}

The local Pauli truncation approximates the original full Heisenberg-evolved operators in Eq.~\eqref{eq:Ot-expansion} using the spatially local truncated operators
\begin{align}
    O_{\tr}(t) = \sum_{\sigma_i \in \bos{P}_{N}^{\tr}}c_i(t) \sigma_i,
\end{align}
where $\bos{P}_{N}^{\tr}$ is the spatially local truncated Pauli string basis with Pauli weight $w$ no larger than a constant $ w^{\tr}$. Therefore, the correlation function
\begin{align}
    C(t):=\bra{\Omega} O(t)O(0)\ket{\Omega}
    \label{eq:Ct-def}
\end{align}
is approximated by the spatially local truncated operator 
\begin{align}
    C_{\tr}(t) :=\bra{\Omega} O_{\tr}(t)O(0)\ket{\Omega}.
\end{align}

In the long-time prediction, the approximation of $C_{\tr}(t)$ using $C(t)$ is feasible because we target the spectral information of $C(t)$ with sparse spectral structure. Because inserting the complete Hamiltonian eigenstates into Eq.~\eqref{eq:Ct-def} gives
\begin{align}
    C(t) =\sum_n |O_{0n}|^2 e^{i(E_0-E_n)t},
\end{align}
where $E_0-E_n$ is the ground-state excitation energy to the $n$th excited state. This excitation spectrum is sparse if we choose operator $O$ with specific quantum numbers, such that only excited states $\ket{E_n}$ with non-zero $O_{0n}=\bra{\Omega}O\ket{E_n}$ contribute to $C(t)$. On the other hand, the Pauli coefficient has spectral information 
\begin{align}
    c_i(t) = \frac{1}{d} \text{Tr} \left[\hat{O}(t)\sigma_i\right]=\frac{1}{d}\sum_{mn} O_{mn}\sigma_{i,nm}e^{i(E_m-E_n)t}.\nonumber
\end{align}
We see that the spectrum $\{E_m-E_n\}$ contains $\{E_0-E_n\}$ with non-zero $O_{0n}$. Therefore, even if we perform the local Pauli truncation, the truncated $O_{\tr}(t)$, and thus $C_{\tr}(t)$, still contain the spectral information of the original $C(t)$. Therefore, we can in principle reconstruct the spectrum of $C(t)$ using $C_{\tr}(t)$.

However, it is possible that the spectrum $\{E_m-E_n\}$ could be too dense to overwhelm the signal $\{E_0-E_n\}$. To test this, we perform numerical studies to examine the effect of local Pauli truncation on the long-time correlation function in both the time and frequency domain. Here, we consider the integrable transverse-field Ising model (TFIM) and the chaotic mixed-field Ising model (MFIM), which correspond to the Ising spin chain
\begin{align}
    H_{\text{Ising}}=-\sum_{i=1}^{N-1} Z_iZ_{i+1}-h\sum_{i=1}^{N} X_i-m\sum_{i=1}^N Z_i
    \label{eq:Ising}
\end{align}
respectively with parameters $(h=1, m=0)$ and $(h=-1.05, m=0.5)$. 

On a spin chain with $N=5$, we plot the truncated correlation function in the time and frequency domain with local Pauli weight $w\leq w^{\tr}\in\{2,3,4\}$ in Fig.~\ref{Fig:LPT-effect}(a) for TFIM and (b) for MFIM. In TFIM, the truncation error is small for all $w^{\tr}\in\{2,3,4\}$. The resulting frequency peaks are all consistent with the full correlation peaks and the spectrum from exact diagonalization (gray solid lines). In comparison, the MFIM correlation function has a larger truncation error, especially for the case $w\leq 2$. The resulting spectrum has more artifactual peaks from $\{E_m-E_n\}$ that are mixed with the target peaks $\{E_0-E_n\}$, making them difficult to distinguish. Nevertheless, despite the presence of artifactual peaks, the frequency spectrum remains dominated by the low-lying target peaks around $\omega=2.8$. Therefore, even when the truncation error in the time domain is substantial, the physically relevant low-lying spectral features remain accessible, and their precision can be further improved using the Neural ODE. The averaged truncation errors over all time points $\overline{|[C_{\tr:w\le w^{\tr}}(t)-C(t)]/C(t)|}$ in TFIM and MFIM are summarized in Fig.~\ref{Fig:LPT-effect}(c), which highlights the smaller truncation error of the integrable TFIM compared with chaotic MFIM.

The physical reason for the larger truncation error of MFIM is the faster spreading of the Heisenberg operator $O_X(t)$ compared with that in TFIM, as we have explained in the main text. This can also be directly observed by the number of discarded Pauli bases, as illustrated in Fig.~\ref{Fig:LPT-effect}(d), where blue and orange bars denote the number of retained Pauli bases for local truncation $w\leq w^{\tr}$. The blank bars denote the number of Pauli bases truncated. For the case without truncation $w\leq 5$, the number of Pauli bases consisting of $O_X(t)$ in TFIM is much smaller than that in MFIM due to the symmetries in TFIM that will be discussed in the next subsection. After the local Pauli truncation, the number of Pauli bases in the chaotic MFIM is much more reduced, leading to a larger truncation error in the resulting correlation function in Fig.~\ref{Fig:LPT-effect}(b) compared with that in Fig.~\ref{Fig:LPT-effect}(a).

\subsection{Pauli reduction by symmetries}
\label{App:symmety}

The operator dynamics can be further reduced by symmetries of the quantum system. We demonstrate this reduction for the transverse-field Ising model (TFIM) studied numerically in the main text. The Hamiltonian of TFIM reads
\begin{align}
H_{\text{T}} = -\sum_{i=1}^{N-1}  Z_iZ_{i+1} - h\sum_{i=1}^N X_i.
\label{eq:TFIM-ham}
\end{align}
This Hamiltonian is invariant under the bit-flip symmetry, i.e., $H_{\text{T}}$ commutes with the global bit-flip operator 
\begin{align}
    [H_{\text{T}}, S] = 0,
    \label{eq:symmetry}
\end{align}
where $S:=\prod_{i=1}^NX_i$.

In this spin system, we show that if an operator $O$ also commutes with $S$, and the Pauli string $\sigma_i$ anti-commutes with $S$, the corresponding Pauli coefficient $c_i(t)$ of $O(t)$ in Eq.~\eqref{eq:Ot-expansion} vanishes. For a Pauli string anti-commuting with $S$, its coefficient reads
\begin{align}
    \begin{aligned}
    c_{i}(t)&=\frac{1}{d}\Tr(e^{iH_{\text{T}}t}SSOe^{-iH_{\text{T}}t}\sigma_i)\\
    &=\frac{1}{d}\Tr(e^{iH_{\text{T}}t}Oe^{-iH_{\text{T}}t}S\sigma_iS)\\
    &=-\frac{1}{d}\Tr(e^{iH_{\text{T}}t}Oe^{-iH_{\text{T}}t}\sigma_i)\\
    &=-c_{i}(t).
\end{aligned}
\end{align}
Thus, we have $c_{i}(t)=0$. In the second line, we use the commutation relation $[O, S]=0$ and Eq.~\eqref{eq:symmetry}. The third line uses the anti-commutation relation $\{\sigma_i, S\}=0$. Since in $\PN$, half of the $\sigma_i$ anti-commute with $S$~\cite{10.1145/3408039}, the bit-flip symmetry of TFIM guarantees that half of the anti-commuting Pauli strings' coefficients vanish among all $4^N$ Pauli bases. This symmetry constraint is imposed a priori in the numerical studies of TFIM by reducing the input and output dimensions $\Ntr$ of the neural networks.

\section{Connection to quantum process tomography}
The operator learning method has a natural interpretation in the language of quantum process tomography (QPT), which also clarifies why a truncated basis is sufficient and why the cost remains polynomial in system size.

\vspace{1.5mm}
\noindent \emph{Full process tomography}~---~The time-evolution map $\mathcal{E}_t$ is in general a completely positive, trace-preserving (CPTP) quantum channel. In the noiseless case, it reduces to the unitary channel $\mathcal{E}_t(\cdot) = e^{-iHt}(\cdot)e^{iHt}$, while in the presence of hardware noise, $\mathcal{E}_t$ acquires a non-unitary component that can damp the Pauli coefficients in time. The process-matrix representation can be used to describe the CPTP map and therefore encompasses both the noiseless and noisy regimes treated in this work. Full QPT reconstructs this channel by representing it in the process matrix with Pauli basis:
\begin{equation}
    [\chi_t]_{ij}
    \;=\; \frac{1}{d}\,\mathrm{Tr}\!\left[\sigma_i\,\mathcal{E}_t(\sigma_j)\right];
    \; \sigma_i,\sigma_j\in\PN.
    \label{eq:chi_full}
\end{equation} 
Complete knowledge of $\chi_t$ is equivalent to complete knowledge of the dynamics, but the matrix has $4^N \times 4^N$ entries, making full QPT exponentially costly in $N$.

\vspace{1.5mm}
\noindent \emph{Targeted tomography of a submatrix}~---~Rather than reconstructing $\chi_t$ in full, we are interested only in predicting a
specific spatially local observable $O$ represented by a linear combination of Pauli strings $O=\sum_l o_l\sigma_l$. Comparing Eq.~(2) in the main text with Eq.~\eqref{eq:chi_full}, the time-dependent coefficient $c_i(t)$ is nothing but a linear combination of the process matrix elements:
\begin{equation}
    c_i(t) = \frac{1}{d}\,\mathrm{Tr}\!\left[O\,\mathcal{E}_t(\sigma_i)\right]=\sum_l o_l [\chi_t]_{l,i}.
    \label{eq:ci_chi_row}
\end{equation}
Learning the dynamics of the Pauli coefficient $c_i(t)$ is therefore equivalent to learning sparse elements of $\chi_t$ in the $i$th column and $\OO(N)$ rows indexed by $l$, rather than the full $4^N \times 4^N$ matrix.
This reduces the output complexity from $\OO(4^N \times 4^N)$ to $\OO(N4^N)$~---~a quadratic improvement with respect to the exponent.

\vspace{1.5mm}
\noindent \emph{Further reduction via local Pauli truncation}~---~The Lieb-Robinson theorem Eq.~\eqref{eq:lieb-robinson-theorem} guarantees that, for finite evolution time $t$, the coefficients $[\chi_t]_{l, i}$ are exponentially suppressed whenever the Pauli string $\sigma_i$ acts on sites that lie outside the effective light cone of $\sigma_l$ with the radius $r\sim vt$. Concretely, for short-time dynamics with $t\leq T$, only spatially local Pauli strings contribute appreciably:
\begin{equation}
    c_i(t) \;\approx\; 0,
    \quad \text{if } \sigma_i \notin \{\text{$2vT$ local Pauli strings}\}.\nonumber
\end{equation}
Therefore, restricting the input index $i$ to this set brings the number of required columns from $4^N$ down to $\OO(4^{k(2vT)^D})$. Combining the locality of $O$ and the local Pauli truncation, the short-time data collect only an $\OO(N4^{k(2vT)^D})$-dimensional submatrix of the full process matrix as given in Eq.~\eqref{eq:Ntr-scaling}.

\begin{figure}
\includegraphics[width=0.48\textwidth]{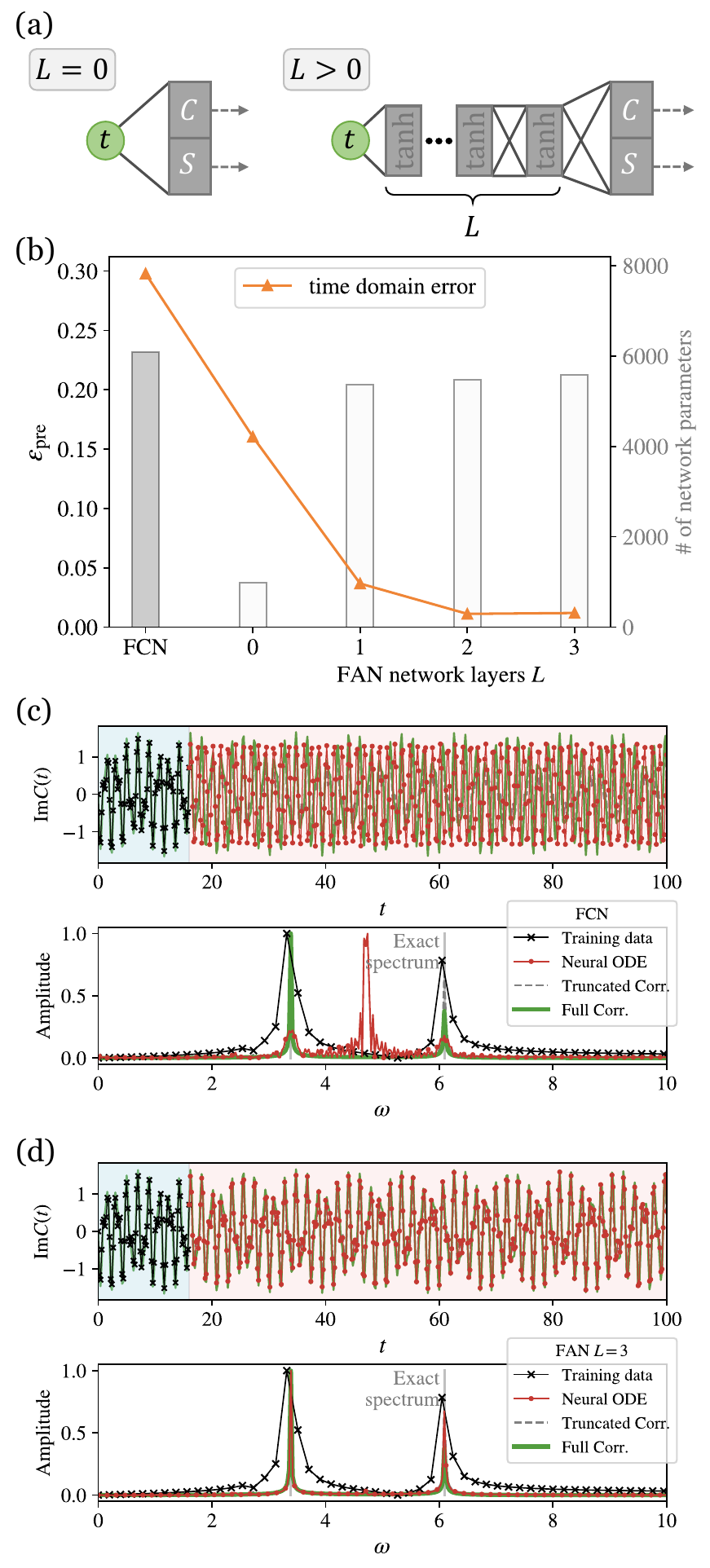}
        \caption{Comparison between the fully connected network (FCN) and the frequency-aware network (FAN) with different envelope network layers $L$. (a) Definition of FAN envelope network layers with $L=0$ and $L>0$. (b) Averaged prediction error using FCN and FAN with $L\in\{0,1,2,3\}$ (orange line marked by upper triangles). The gray bars denote the total number of training parameters in these networks. Imaginary part of correlation functions predicted by (c) FCN and (d) $L=3$ FAN in the time (upper panel) and frequency domain (lower panel). The prediction results are denoted by red-dotted lines. The training data are black-cross lines with the number of time points $N_d=81$.}
    \label{fig:NNcompare}
\end{figure}

\section{Prediction errors and ablation studies} \label{app:ablation-studies}

To investigate the Neural ODE prediction errors of the operator learning method, we examine the frequency-aware network (FAN) components and compare FAN with a fully connected network (FCN). Additionally, we also study the impact of different data sizes, symmetry constraints, and Hamiltonian coupling strength on the resulting prediction errors. To quantify the prediction error, we define the following averaged prediction error:
\begin{align}
    \varepsilon_{\text{pre}} := \frac{1}{N_\text{pre}}\sum_{t_j\in\{\text{pre}\}}\left|\frac{C_{\text{pre}}(t_j)-C_{\tr}(t_j)}{C_{\tr}(t_j)}\right|,
\end{align}
where $\{\text{pre}\}$ is a set of the Neural ODE prediction time points with $\delta t=0.2$. $N_\text{pre}$ is the total number of prediction time points. $C_{\text{pre}}(t_j)$ and $C_{\tr}(t_j)$ are respectively the correlation functions $\bra{\Omega} O_{X,\tr}(t)O_X(0)\ket{\Omega}$ derived from the Neural ODE prediction and the ground truth results, where the ground truth results are derived by the $w\leq 2$ local Pauli truncation of an $N=3$ TFIM spin chain.

\vspace{1.5mm}
\noindent \emph{FAN v.s. FCN and FAN layers}~---~We first compare the prediction error of FAN and FCN defined in the main text, and investigate the effect of FAN envelope network layers on the results. The envelop network layers $L=0$ and $L>0$ are illustrated in Fig.~\ref{fig:NNcompare}(a), where the $L=0$ case is a biased linear transformation 
\begin{equation}
    \begin{aligned} 
\begin{bmatrix}
C_{ijm}\\
S_{ijm}
\end{bmatrix}&=\mathbf{W}^{(e)}_{ijm}\times t+\mathbf{b}_{ijm}^{(e)},
\end{aligned}
\end{equation}
and $C_{ijm}=C_{ijm}(t), S_{ijm}=S_{ijm}(t)$ are input to $\hat{G}_{\theta}(t)$ as in Eq.~(4) of the main text. The averaged prediction errors of FCN and FAN with envelope layers $L\in[0,3]$ are shown in Fig.~\ref{fig:NNcompare}(b), and the corresponding number of network parameters is plotted using gray bars. We see that the FCN has a larger prediction error than FAN even using more trainable parameters. For FAN, increasing the number of network layers leads to smaller prediction errors, demonstrating the importance of neural network depth for accurate long-time prediction. The Imaginary part of $C(t)$ in the time and frequency domain predicted using FCN and FAN with $L=3$ is plotted in Fig.~\ref{fig:NNcompare}(c) and (d), respectively. We see that the FAN with $L=3$ predicts the frequency peaks consistent with the exact spectrum, whereas the FCN predicts an incorrect frequency peak.

\begin{figure}
\centering
\includegraphics[width=0.48\textwidth]{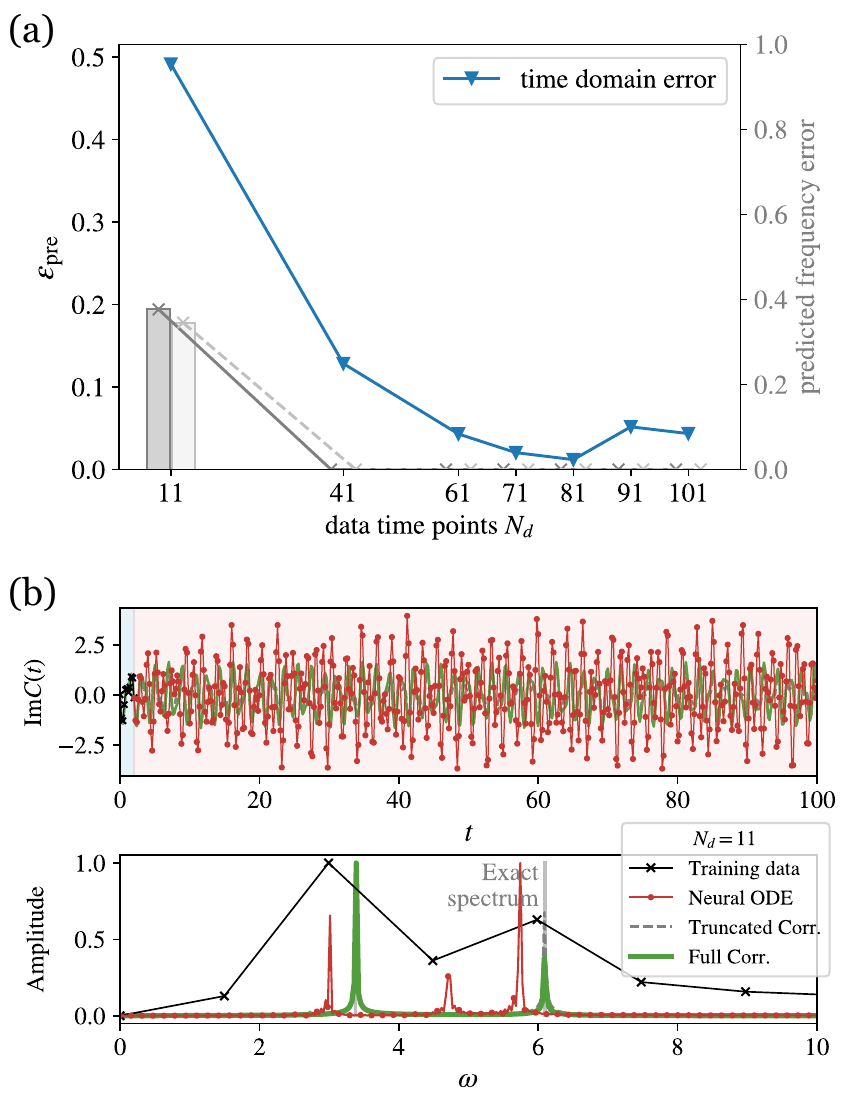}
        \caption{Impact of different data sizes. (a) Averaged prediction error versus the number of data time points $N_d$ (blue line marked by lower triangles). The gray bars denote the error of two predicted frequency peaks for different $N_d$. (b) Imaginary part of the correlation function predicted by the number of data points $N_d=11$. An insufficient number of data points leads to incorrect frequency peaks, as shown explicitly in the lower panel of (b) and the non-zero gray bars in (a). }
    \label{fig:data-size}
\end{figure}
\begin{figure}
\centering
\includegraphics[width=0.49\textwidth]{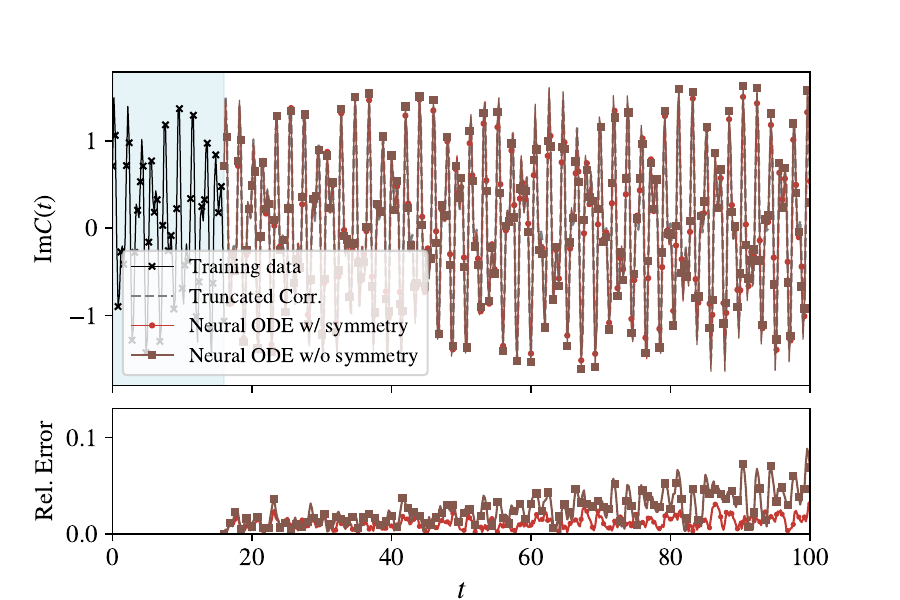}
        \caption{Neural ODE prediction of the correlation function with (red-dotted line) and without (brown-square line) symmetry constraints. The relative error in the lower panel is derived by comparing the two predictions with the ground truth results with $w\leq 2$ local Pauli truncation.}
    \label{fig:symmetry}
\end{figure}

\vspace{1.5mm}
\noindent \emph{Data size}~---~Using FAN layers $L=3$, we study the impact of different data sizes on the prediction accuracy. Fig.~\ref{fig:data-size}(a) plots the averaged prediction error versus the number of data time points $N_d$ used to train FAN. As $N_d$ increases, the prediction error decreases until $N_d = 81$. The error increases at $N_d > 81$ mainly due to insufficient training iterations. The prediction error is also reflected in the frequency domain, which is shown by the gray bars in Fig.~\ref{fig:data-size}(a). The predicted frequency at $N_d=11$ deviate from the exact spectrum, as explicitly shown in the lower panel of Fig.~\ref{fig:data-size}(b). In contrast, increasing the training data to $N_d=81$ gives more accurate predictions in both the time and frequency domain, as we have seen in Fig.~\ref{fig:NNcompare}(d). These results highlight the essential role of a large enough data size for accurate long-time dynamics prediction.

\vspace{1.5mm}
\noindent \emph{Symmetry constraints}~---~Symmetry constraints on the Pauli coefficients $c_{i}(t)$ are exact without involving additional errors. The constraints lead to certain coefficients $c_{i}(t)=0$ that are not necessarily involved in the training process, as discussed in Sec.~\ref{App:symmety}. Discarding these vanished $c_{i}(t)$ a priori improves the training efficiency and avoids possible overfitting. Fig.~\ref{fig:symmetry} plots the Neural ODE prediction of the correlation function of the $N=3$ TFIM with and without symmetry constraints. In this example, the symmetry reduces the input dimension $\Ntr$ from 27 to 11, and the number of FAN parameters is reduced from $32347$ to $5579$. With these reductions, as shown in the lower panel of Fig.~\ref{fig:symmetry}, the Neural ODE prediction with symmetry has a smaller relative error compared with the one without symmetry, especially at longer prediction time $t$. These results showcase that symmetry constraints can improve the long-time prediction accuracy.

\begin{figure}
\centering
\includegraphics[width=0.48\textwidth]{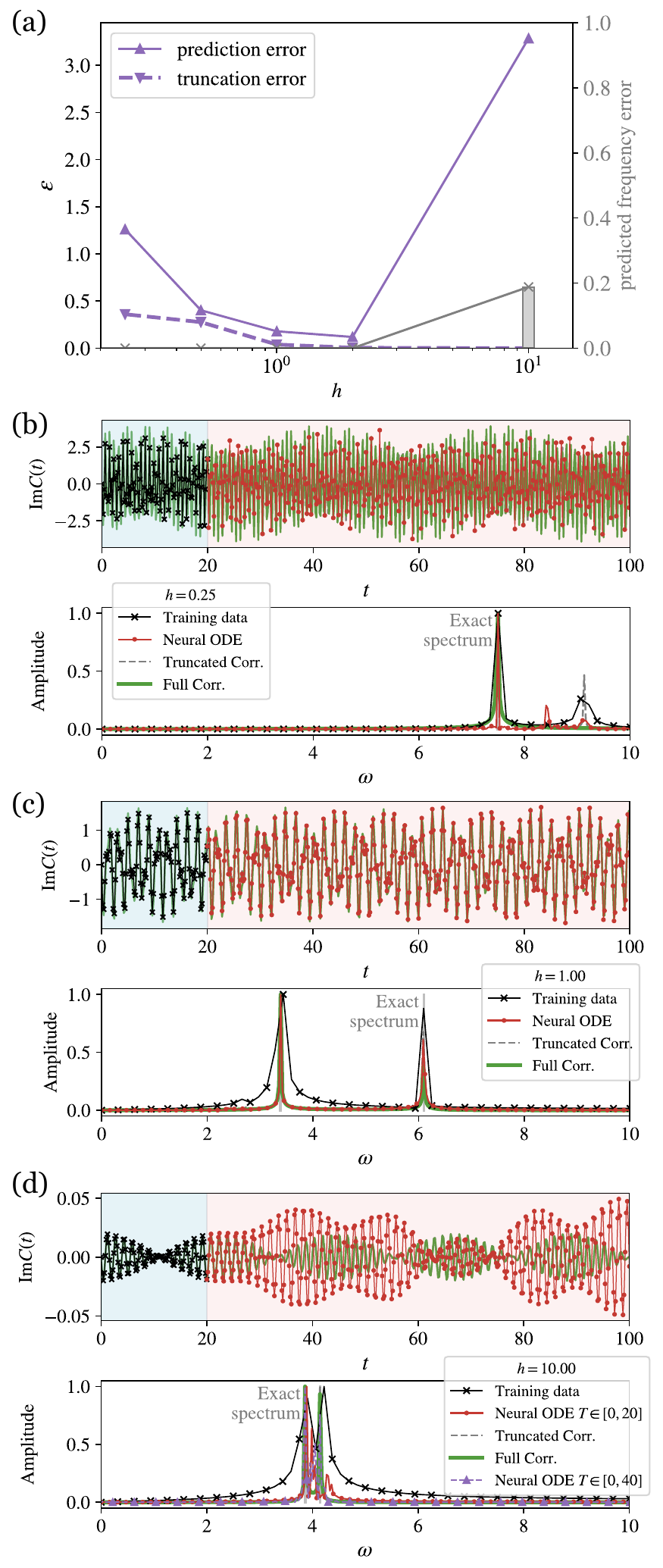}
        \caption{Impact of the external field strength $h$ in TFIM. (a) Averaged prediction error and local Pauli truncation error in the time domain for different $h$. The corresponding frequency errors are plotted using gray bars. The x-axis is in the log scale. Imaginary part of the correlation functions for (b) $h=0.25$, (c) $h=1.0$, and (d) $h=10.0$. For $h=10.0$, the Neural ODE prediction using data time length $T\in[0,40]$ is marked using upper triangles, which gives correct frequency peaks.}
    \label{fig:h-dependence}
\end{figure}

\vspace{1.5mm}
\noindent \emph{External field strength}~---~We study the prediction error dependence on the external field strength $h$ in the TFIM Hamiltonian defined in Eq.~\eqref{eq:TFIM-ham}. Here we use FAN with $L=3$ and data time points $N_d=101$. Fig.~\ref{fig:h-dependence}(a) plots the averaged prediction error versus $h$, as well as the averaged local Pauli truncation error
\begin{align}
    \varepsilon_{\tr} := \frac{1}{N_\text{pre}}\sum_{t_j\in\{\text{pre}\}}\left|\frac{C_{\tr :w\leq2}(t_j)-C(t_j)}{C(t_j)}\right|,
\end{align}
and the predicted frequency error. Both the prediction error and truncation error are large at small $h$ because a small $h$ corresponds to a large $ZZ$ coupling strength, such that the coupling strength between the local Pauli strings and the discarded non-local ones is large. Specifically, $L_{PQ}$ of the memory kernel spectral decomposition in Eq.~(16) of the main text is large. However, despite the large error in the time domain, the frequency can still be predicted accurately at small $h$. This is explicitly shown by the correlation function spectrum in the lower panel of Fig.~\ref{fig:h-dependence}(b) with $h=0.25$, where the frequency peak at around $\omega = 7.5$ is consistent with the spectrum by exact diagonalization. This accuracy difference in the time and frequency domains has been observed in Fig.~\ref{Fig:LPT-effect}, which is because the truncated correlation function still contains the spectral information of the full correlation. By increasing $h$ from a small value, both the prediction error and truncation error decrease, as shown in Fig.~\ref{fig:h-dependence}(c) with $h=1$ as an example. 

However, further increasing $h$ again leads to a large prediction error, and the predicted frequencies are also incorrect, as shown by the gray bars in Fig.~\ref{fig:h-dependence}(a). This is because the TFIM spectrum at large $h$ tends to degenerate, as illustrated by the two neighboring peaks in Fig.~\ref{fig:h-dependence}(d). The FAN is hard to learn different frequencies if they are not clearly distinguished in the training data. In contrast, increasing the training data time from $T\in[0,20]$ to $T\in[0,40]$ allows the data and thus FAN to distinguish the two peaks and recover the correct frequency peaks, as plotted by the purple upper-triangle line in Fig.~\ref{fig:h-dependence}(d).

In summary, the numerical results in this section verify several regimes in which the prediction accuracy of the operator learning method degrades. Inadequate neural-network depth, insufficient training data, fast operator spreading, and overly dense spectrum can all degrade the accuracy of the predicted time-domain signal. This can happen in strong spatial coupling and non-integrable Hamiltonian systems, and highly excited states or critical systems with dense spectra. Nevertheless, in most cases, correct spectral information can still be extracted from the predictions. Moreover, the prediction accuracy can be systematically improved by increasing the amount of training data and neural network layers.
\begin{figure*}
\centering
\includegraphics[width=1\textwidth]{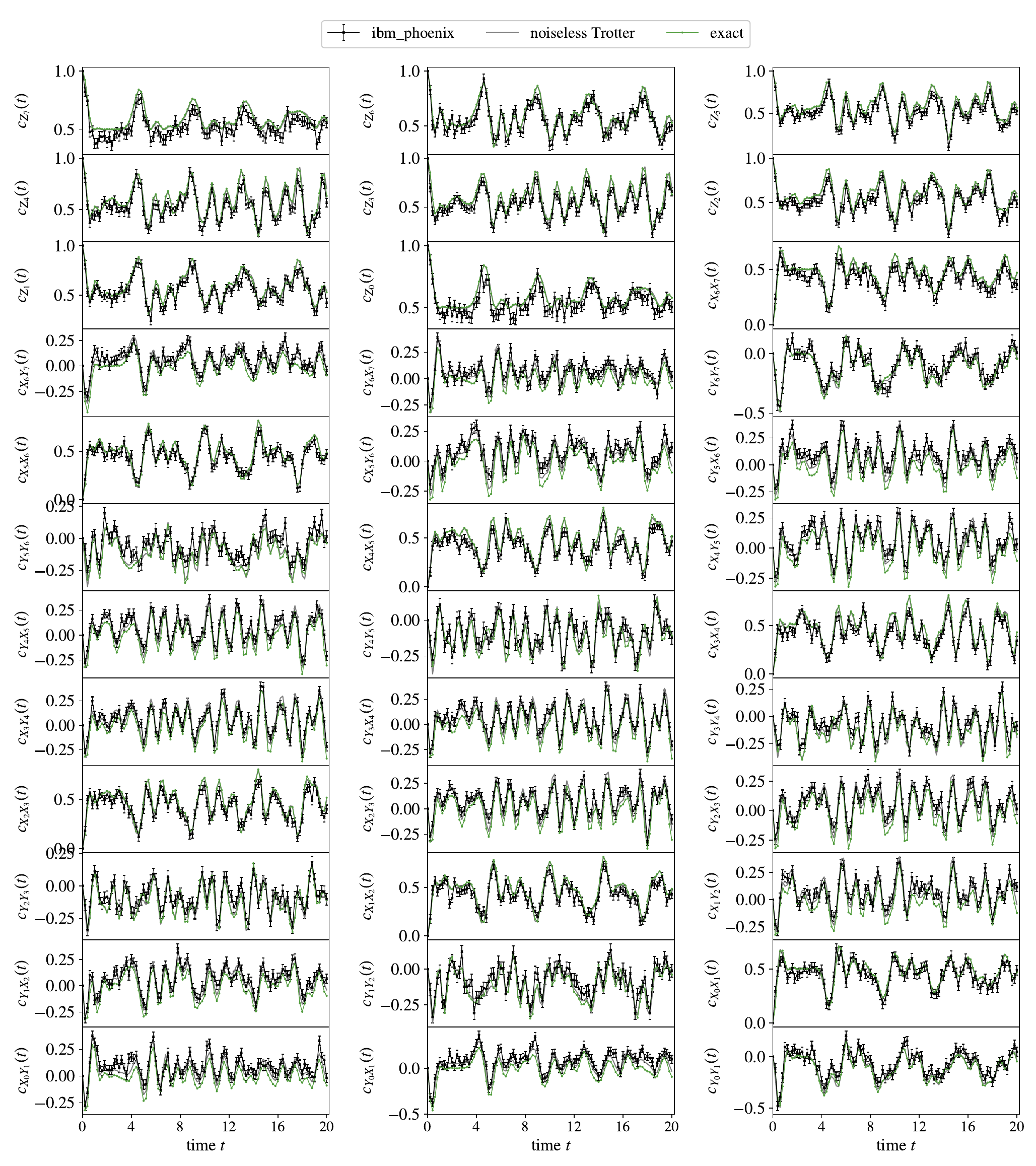}
\cprotect\caption{Measurement data of the 36 Pauli coefficients $c_{i}(t)=\Tr(O_X(t)\sigma_i)$ with local weight truncation $w\leq 2$. Results from hardware \verb|ibm_pheonix|, noiseless Trotter, and exact diagonalization are plotted using black-cross, gray, and green-dotted lines. Each Pauli coefficient is measured for 100 time points $t\in{0.2,0.4,\ldots,20}$. Therefore, $36\times 100=3600$ circuits are executed in total. For each circuit corresponding to one data point, $8192$ shots are sampled to acquire statistics.}
\label{fig:Pauli_coeff}
\end{figure*}
\section{Spectroscopy of two-point correlation functions}
\label{App:spectrum}

In this section, we present how the two-point correlation function can be used to study the spectroscopy of quantum many-body systems. The two-point correlation function in terms of its Fourier transform reads
\begin{align}
C(t) = \langle \Omega| O(t) O(0) |\Omega\rangle = \int_{-\infty}^{\infty} \frac{d\omega}{2\pi}, e^{-i\omega t} S(\omega).
\end{align}
Here, $|\Omega\rangle=\ket{E_0}$ is the ground state of the quantum system. $S(\omega)$ is the spectral function of the quantum many-body system, which is given by performing the inverse Fourier transform of $C(t)$
\begin{equation}
\begin{aligned}
S(\omega) &= \int_{-\infty}^{\infty} dt\, e^{i\omega t} C(t) \\
&= \int_{-\infty}^{\infty} dt \sum_{n} e^{i(\omega - (E_n-E_0))t} |\langle E_0 | O(0) | E_n \rangle|^2  \\
&= 2\pi \sum_{n} |\langle E_0 | O(0) | E_n \rangle|^2 \delta(\omega - (E_n-E_0)).
\end{aligned}
\end{equation}
In the second line, we insert the complete set of energy eigenstates $\sum_{n} |E_n\rangle \langle E_n| = \mathbb{I}$ between $O(t)$ and $O(0)$ of the correlation function. The $\delta$ function in the third line shows that the frequency value where $S(\omega)$ is peaked corresponds to the excitation energy $E_n-E_0$ of the target Hamiltonian.

In the numerical results in this work, the correlation function of the Ising spin chain with mixed fields can be doubled in the negative time direction according to the relation
\begin{align}
C^*(t) = C(-t),
\label{eq:double-time-relation}
\end{align}
where $C^*(t)$ is the complex conjugation of $C(t)$. This relation holds because the Ising spin chain has time-reversal symmetry, i.e., its Hamiltonian in Eq.~\eqref{eq:Ising} is invariant under complex conjugation on each of its matrix elements
\begin{align}
H_{\text{Ising}} = H_{\text{Ising}}^*,
\end{align}
and the ground state $\ket{\Omega}$ also has all real element $\ket{\Omega} = \ket{\Omega}^{}$. Using this time-reversal symmetry, the complex conjugation of $C(t)$ reads
\begin{equation}
\begin{aligned}
C^*(t) &= \bra{\Omega}^* (e^{iH_{\text{Ising}}t})^* (O_X)^* (e^{-iH_{\text{Ising}}t})^* (O_X)^*\ket{\Omega}^*\\
&=\bra{\Omega} e^{-iH_{\text{Ising}}^*t} O_X e^{iH_{\text{Ising}}^*t}O_X\ket{\Omega}\\
&=\bra{\Omega} e^{-iH_{\text{Ising}}t} O_X e^{iH_{\text{Ising}}t}O_X\ket{\Omega}\\
&=C(-t).\nonumber
\end{aligned}
\end{equation}
Therefore, we can double the measured $C(t)$ in the negative time direction. Eq.~\eqref{eq:double-time-relation} also explains that the imaginary part of $C(t)$ plotted in this work always has $\text{Im}C(0)=0$.

\begin{figure}
\centering
\includegraphics[width=0.47\textwidth]{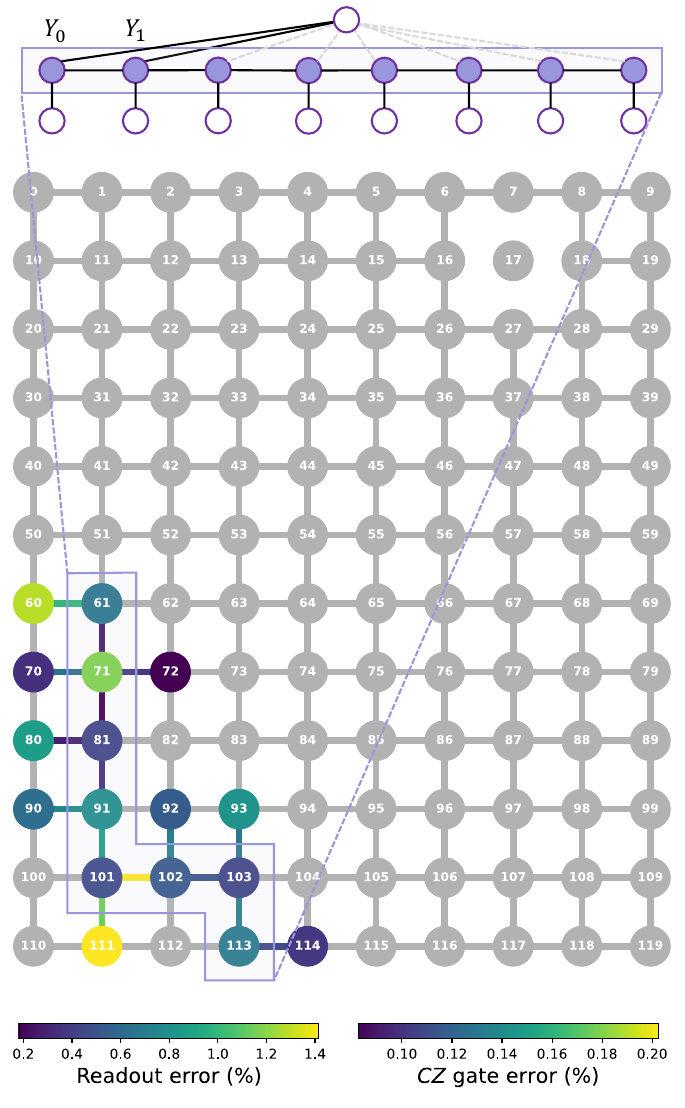}
\cprotect\caption{Qubit layout and mapping onto \verb|ibm_phoenix|'s coupling map to measure the Pauli coefficient $c_{Y_0Y_1}(t)$. The coupling map is colored to represent the readout error for a single qubit and the two-qubit $CZ$ gate error for each qubit connection.}
\label{fig:layout}
\end{figure}

\section{Hardware data and implementation details}
\label{app:FAN_FCN}
In the hardware implementation for predicting the correlation function of the $N=8$ TFIM, the FAN is trained on coefficients $c_i(t)$ of 36 Pauli bases $\sigma_i$ with local weight $w\leq 2$ and $N_d=101$ data time points. These 36 data trajectories measured using \verb|ibm_phoenix| are presented in Fig.~\ref{fig:Pauli_coeff}. The results from hardware, noiseless Trotter simulation, and exact diagonalization are plotted using black-cross, gray, and green-dotted lines. Each data point is obtained from 8192 shots to acquire statistics, and the black error bar denotes the statistical error. During the execution of the quantum circuits, we use dynamic decoupling~\cite{Ezzel2023} and randomized compiling~\cite{PhysRevX.11.041039} to suppress the time-keeping and coherent errors. We see that the errors from Trotter decomposition and hardware noise have the same magnitude in most trajectories. However, despite the existence of these Trotter errors, hardware noise, and statistical error, the oscillatory behavior of the Pauli coefficients is clearly observed and well consistent with the exact results, ensuring that the FAN can learn the periodic modes using these short-time training data.

The quantum circuits are executed on 17 qubits on the 120-qubit device \verb|ibm_phoenix|. We use the Qiskit \verb|optimization_level=3| to map the 17 qubits onto the 17 physical qubits of \verb|ibm_phoenix| with noise as weak as possible. For example, Fig.~\ref{fig:layout} shows the layout and mapping onto \verb|ibm_phoenix|'s coupling map to measure $c_{Y_0Y_1}(t)$, plotted in the bottom-right subplot of Fig.~\ref{fig:Pauli_coeff}. The coupling map is colored to represent the readout error rate for each qubit and the two-qubit $CZ$ gate error rate for each qubit connection. These hardware error rate data are obtained from the IBM cloud quantum platform~\cite{Qiskit}.

\begin{figure}[b]
\centering
\includegraphics[width=0.48\textwidth]{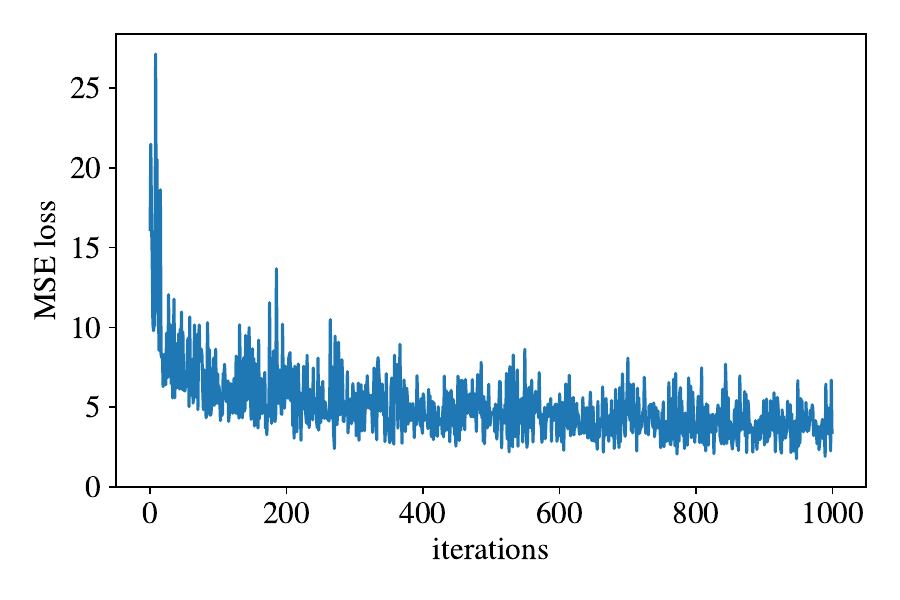}
\caption{MSE loss as a function of iterations in the 8-qubit training process using the hardware data. The training is performed using the Adam optimizer~\cite{kingma2017adammethodstochasticoptimization} and stopped after 1000 iterations.}
\label{fig:MSE-loss}
\end{figure}

\section{Neural ODE training details} \label{app:Neural ODE}

In the training and prediction process of Neural ODE, we use the fourth-order Runge-Kutta ODE initial value solver with step length $0.005$, which ensures stable evolution of the Pauli coefficients up to $t=100$. For the 3-qubit Ising spin system simulation, the FAN with layer $L=3$ and hidden dimension $d_h=10$ has a number of trainable parameters of $\OO(10^3)$, and the $8$-qubit hardware implementation has 199,909 parameters.

The loss function used to train the FAN is defined by the mean-squared error (MSE) between the data and predicted coefficients,
\begin{align}
\mathcal{L} = \frac{1}{\Ntr\times |{t_j}|} \sum_{\sigma_i\in\PNtr}\sum_{{t_j}} \left| c_i^{\text{data}}(t_j) - c_i^{\text{pre}}(t_j) \right|^2,
\end{align}
where the training time points ${t_j}$ are obtained as follows~\cite{chen2018neuralode,3454287.3454765}: Given a set of time points and a maximum sample size to be half the data time points $N_{\max}=\lfloor N_d/2\rfloor$, we first uniformly sample an initial time point $t_1$. If the number of available time points from $t_1$ onward does not exceed $N_{\max}$, all of them are included in the sample. Otherwise, we randomly select $N_{\max}-1$ points from those after $t_1$ and combine them with $t_1$, yielding a sample of size $N_{\max}$. The training is performed using the Adam optimizer~\cite{kingma2017adammethodstochasticoptimization} with a learning rate of 0.001 and up to 1000 iterations. With these training parameter settings, the training converges for all numerical results presented in this work. For example, Fig.~\ref{fig:MSE-loss} shows the MSE loss as a function of iterations in the 8-qubit training process using the hardware data. The MSE loss decreases steadily and shows good convergence after approximately 1000 training iterations.
\end{document}